\documentclass[sigconf]{acmart}

\usepackage[ruled]{algorithm2e}
\usepackage{subcaption}
\usepackage{caption}
\usepackage{listings,multicol}
\usepackage{enumitem}
\usepackage{todonotes}
\usepackage[htt]{hyphenat}
\usepackage{soul}
\usepackage{dblfloatfix}

\usepackage{fancyhdr}

\copyrightyear{2024}
\acmYear{2024}
\acmDOI{XXXXXXX.XXXXXXX}

\acmConference[PASC '24]{the Platform for Advanced Scientific Computing Conference}{June 03--05, 2024}{Zürich, Switzerland}

\acmPrice{XX.XX}
\acmISBN{978-1-4503-XXXX-X/18/06}

\begin{document}

\fancyhead[]{}
\fancyhead[CO,CE]{ACCEPTED AT THE ACM PLAFORM FOR ADVANCED SCIENTIFIC COMPUTING (PASC) CONFERENCE, 2024}
\title{Reducing the Impact of I/O Contention in Numerical Weather Prediction Workflows at Scale Using DAOS}

\author{Nicolau Manubens}
\email{nicolau.manubens@ecmwf.int}
\affiliation{%
  \institution{European Centre For Medium-Range Weather Forecasts}
  \country{Germany}
}

\author{Simon Smart}
\email{simon.smart@ecmwf.int}
\affiliation{%
  \institution{European Centre For Medium-Range Weather Forecasts}
  \country{United Kingdom}
}

\author{Emanuele Danovaro}
\email{emanuele.danovaro@ecmwf.int}
\affiliation{%
  \institution{European Centre For Medium-Range Weather Forecasts}
  \country{United Kingdom}
}

\author{Tiago Quintino}
\email{tiago.quintino@ecmwf.int}
\affiliation{%
  \institution{European Centre For Medium-Range Weather Forecasts}
  \country{United Kingdom}
}

\author{Adrian Jackson}
\email{a.jackson@epcc.ed.ac.uk}
\affiliation{%
  \institution{The University of Edinburgh}
  \country{United Kingdom}}

\renewcommand{\shortauthors}{Manubens et al.}

\begin{abstract}
Operational Numerical Weather Prediction (NWP) workflows are highly data-intensive. Data volumes have increased by many orders of magnitude over the last 40 years, and are expected to continue to do so, especially given the upcoming adoption of Machine Learning in forecast processes. Parallel POSIX-compliant file systems have been the dominant paradigm in data storage and exchange in HPC workflows for many years. This paper presents ECMWF's move beyond the POSIX paradigm, implementing a backend for their storage library to support DAOS --- a novel high-performance object store designed for massively distributed Non-Volatile Memory. This system is demonstrated to be able to outperform the highly mature and optimised POSIX backend when used under high load and contention, as per typical forecast workflow I/O patterns. This work constitutes a significant step forward, beyond the performance constraints imposed by POSIX semantics.
\end{abstract}





\maketitle

\thispagestyle{fancy}

\section{Introduction}

Numerical Weather Prediction (NWP) uses mathematical models to predict the conditions of the atmosphere and other components of the Earth system over varying forecast horizons, from hours to weeks in advance. These models include direct physical simulations and Machine Learning (ML) models\cite{ecmwf-forecasts}. Due to the diverse and vast quantities of data ingested and produced by these models, as well as the complexity of NWP workflows, NWP model runs are considered characteristic data-intensive applications. Data volumes will continue to increase substantially in the coming years, as scientific ambition drives higher model resolution and larger numbers of perturbed simulation replicas\cite{forecast-skill-bauer}, and the range of downstream applications consuming forecast data is also commensurately growing. This poses a significant storage capacity requirement for current and future NWP data centres.

The challenge is further magnified by the time-critical nature of forecasting systems, driven by the sharply decreasing value of a forecast as time passes. Applications producing and consuming data run simultaneously in the HPC system, and data needs to be exchanged efficiently even under significant load and system contention.

Large distributed POSIX file systems backed by hard-drive disks (HDD) and solid-state disks (SSD) are currently the predominant high-per\-formance storage solution in HPC systems. These file systems present limitations in highly parallel data-inten\-sive workloads. On one hand, the POSIX file system API forces the application developer to use files and directories as the data storage units, and to think about how to distribute the data in them. The developer needs to consider concurrent access to the files and directories from several parallel processes, and implement mechanisms to ensure consistency while maintaining a sufficient level of performance. This becomes not only a source of complexity and errors in the application, but also an extremely challenging task even for an experienced programmer\cite{lustre-practices}. If multiple applications need to access the same data in the file system concurrently, all of the applications should include mechanisms to address these issues, which can be complex and costly.

On the other hand, the APIs and semantics exposed by POSIX file systems were designed decades ago when storage devices were attached only locally to the machine rather than over the network, and the POSIX standards\cite{posix-standard} have inherited and mandate features that can hinder performance in highly-parallel workloads. Specifically, POSIX prescribes lots of metadata to be maintained by the operating system for each file and directory; the consistency guarantees are sometimes excessive\cite{lockwood}; it relies on the operating system's block device interface, which enforces retrieval of entire blocks even if only a few bytes in a file are requested; and it mandates file semantics that 
are over-constrained and non-optimal for high write and read contention workloads on distributed systems, such that distributed locking mechanisms need to be put in place by the distributed file system implementations, causing large lock communication overheads on the client nodes\cite{io-contention-filesystems}\cite{gpfs-internals}\cite{lustre-internals}.

The European Centre for Medium-Range Weather Forecasts (ECMWF) developed a storage library, named the FDB\cite{fdb-pasc2017}\cite{fdb-github}, which exposes a domain-specific API --- described in in section\ \ref{sect:fdb} --- for NWP applications to store and index weather fields according to scientifically meaningful metadata. This was built with a software backend making use of a POSIX-compliant distributed file system, handling the use and parallel access to files and directories behind the scenes and thus decoupling as best as possible the behavioural semantics required by the application domain from those provided by the file system.

More recent work has functionally split the FDB POSIX backend into \emph{Catalogue} and \emph{Store} backends which implement the indexing functionality and data storage capability, respectively.

Notwithstanding this work in FDB to address the challenges associated with using and managing parallel access to files and directories, the other issues associated with POSIX APIs and semantics have remained largely unaddressed for years.

Currently, the operational storage systems at ECMWF are already being used at their performance limit, and expanding them to support upcoming data volume increases of one or two orders of magnitude would be costly, or even not viable due to the unprecedented degree of contention. With the goal of enabling operational NWP to scale to such larger data volumes, ECMWF conducted a preliminary assessment\cite{daos-ipdps} of a new storage technology called Distributed Asynchronous Object Store (DAOS)\cite{daos-scfa2022}, which is a high-performance object store designed for massively distributed Non-Volatile Memory (NVM), originally developed by Intel and recently transferred to the newly formed DAOS foundation. DAOS not only offers an object store API with novel semantics which allow for server-side contention resolution, but it also reduces metadata requirements, operates fully in user-space and allows for byte-granular access to NVM storage devices. The preliminary assessment yielded promising results, but it was conducted by developing a standalone benchmark, independent from ECMWF's operational software storage stack.

This paper presents the development and evaluation of DAOS Catalogue and Store backends for the FDB. These backends were benchmarked against DAOS deployments on the NEXTGenIO prototype system\cite{ngio} containing NVM distributed across the nodes. A performance comparison against identical workflows using the POSIX backends and Lustre on the same hardware is included.

\subsection{Related work}

Hardware and software storage technologies have proliferated in the last decade\cite{storage-survey}. There have been many research works on how to exploit and utilise these on the path to adapting I/O to Exascale. Some have focused on the benefits of object storage as opposed to commonly used file systems and their limitations\cite{object-storage-perf}\cite{ceph-adopting-object-storage}. Others have focused on adapting I/O middleware to exploit object storage\cite{hdf5-adopting-object-storage}, and some have reported successful outcomes with DAOS specifically as a backend for I/O middleware\cite{rexio-daos-interfaces}\cite{tpds-hdf5}\cite{pdsw23-adios} or file system APIs\cite{dfs-performance} on top of DAOS.

Object stores are increasingly present in IO-500 performance rankings\cite{io500-sc23}, with some large institutions at the top after adopting object storage --- mostly as a backend for the file system APIs commonly used in their applications.

Some have gone beyond file system and I/O middleware APIs and developed domain-specific object stores building on top of general-purpose object stores and low-level NVM storage libraries, and verified the benefits of that approach\cite{cray-domain-specific-os}. This includes ECMWF's previous work on the FDB5\cite{fdb-pasc2017}, which originally exploited POSIX file systems, and later was adapted to make native use of PMEM\cite{fdb-pasc2019} and the Ceph RADOS and Cortx Motr object stores.
Part of the conclusions of this work were that a PMEM backend was difficult to maintain, and RADOS, which is currently being reevaluated, did not at the time provide sufficient capabilities to implement FDB indexing. Of the remaining storage options, Motr was recently discontinued.

ECMWF has recently assessed DAOS as a potential backend for their FDB. First, a performance analysis was conducted\cite{daos-ipdps} with standalone benchmark applications making native use of DAOS (Field I/O and IOR\cite{ior} with its DAOS API) showing very good performance. A dummy DAOS library was later developed which maps the DAOS API onto a file system API, and the benchmarks were run again using that library, thus exploiting Lustre deployments under the hood. DAOS showed much better performance if compared to dummy DAOS on Lustre\cite{daos-pdsw}. That comparison, however, was not perfect because runs with dummy DAOS abused the file system API with massive amounts of small I/O operations rather than following file system best practices.

The work presented here describes a newly developed FDB backend which enables native FDB operation on DAOS, and discusses the implementation and performance differences between running realistic NWP I/O workflows with the new FDB backend on DAOS vis-à-vis the traditional POSIX backend on Lustre --- the latter making careful and efficient use of the file system API.

\subsection{Numerical Weather Prediction Data Flows}
\label{sect:nwp}

At ECMWF, the operational forecasting system is run 4 times a day in 1-hour time-critical windows. During each of these windows an ensemble of 52 perturbed model instances (members) are run across approximately 2500 compute nodes. For each of the 240 model output steps, global two-dimensional slices of data (fields) are aggregated by 2500 I/O server processes running on dedicated I/O nodes and written into a Lustre distributed file system with the FDB. Approximately 70 TiB of data are produced by the models and stored by the I/O servers during an operational run, comprising approximately 25 million fields. 70\% of the data is immediately consumed by an ensemble of post-processing tasks which run alongside the models generating derived products.

\begin{figure}[htbp]
    \includegraphics[width=240pt]{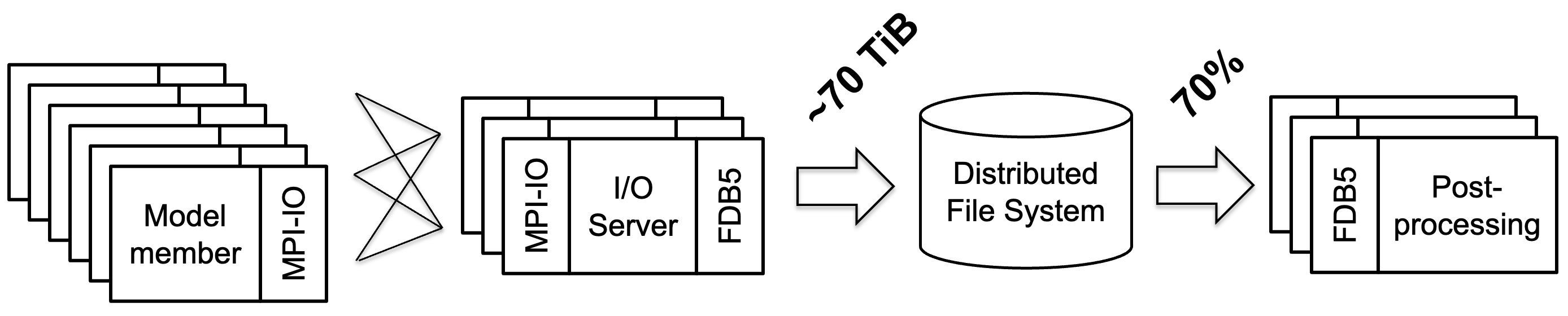}
    \caption{A simplified overview of ECMWF's operational work flow}
\end{figure}
\label{fig:ecmwf_data_flow}

Downstream consuming processes need to be launched as soon as their input data has been made available. Ensuring that the model, and thus the workflow controller, knows when data has been made available drives one of the primary semantic requirements of the FDB. Accomplishing this in the operational workload, under the contention of thousands of simultaneous reading and writing processes provides one of the most significant technical challenges for the FDB.

This is exacerbated by the nature of the write and read patterns. On one hand, each of the I/O server processes writes a stream of fields, which belong to a single member of the simulation. These streams contain fields for a range of simulated variables and levels in the atmosphere, with the processes producing and streaming fields for a sequence of time steps as the model progresses.

On the other hand, the post-processing tasks that consume the data are largely specified per step. That is, once step $n$ of the forecast is completed, the post-processing for step $n$ is launched, and those processes read data as a slice across all the output streams, but only for step $n$. This is essentially a transposition of the view of the data from that of the I/O server, and requires \emph{each} of the post-processing processes to interact with \emph{every} indexing and data output stream produced by the I/O server processes, whilst the model continues to produce and stream data for the following steps.

How this contention is handled in detail is covered extensively in an earlier paper\cite{fdb-pasc2017}. But fundamentally, in the POSIX backends, the write pathway is optimised to the benefit of the writing processes. Each process writes its own independent indexing and data files, and maintains the transactionality required of the FDB by careful insertion of entries on the end of a table of contents file, making use of the precise semantics of the \verb!O_APPEND! mode of a POSIX-compliant file system. The levels of indirection required to make the write pathway performant come at a cost to the read pathway which must access many table of contents and index files to locate the data. This read pathway has been aggressively optimised to be \emph{good enough} for the operational purposes, with extensive index preloading, caching and pruning of the index exploration process.

A consequence of this design is that in a system with minimal contention, the write pathway is expected to perform close to the limits of the underlying file system, and the read pathway is expected to be a little slower. Once runtime contention is introduced between read and write processes, the performance impact of the internal file system lock exchanges between the storage servers and the clients required to maintain consistency is expected to be noticeable. Within DAOS, the distribution of metadata operations across all servers, server-side handling of consistency, and the ability to undertake fine-grained I/O operations, are anticipated to significantly reduce the observed cost of this contention, leading to a benefit on a highly-contended system when running at scale.

\subsection{The FDB}
\label{sect:fdb}

The FDB is a domain-specific object store for meteorological data, which sits between various data producing and consuming components in wider NWP workflows. In practice, it is a library which is compiled into these components, along with a set of management command-line tools, and configurations and data governance rules to establish a particular usage pattern. This library exists to abstract away the specific behavioural details of various underlying storage systems, and to provide a standardised API for data use.

The FDB API is metadata-driven --- that is, all API actions are invoked using scientifically-meaningful metadata describing the data to be acted upon. All data objects (fields) are identified by a globally-unique metadata identifier, formed as a set of key-value pairs conforming to a user-defined schema. An example field identifier is illustrated in Listing\ \ref{listing:mars-request}.

\begin{multicols}{3}[\captionof{lstlisting}{An example metadata identifier of a field in FDB.}]
\begin{lstlisting}[label=listing:mars-request,xleftmargin=0.0cm]
class = od,
date = 20231201,
levtype = sfc,
step = 1,
stream = oper,
time = 1200,
number = 13,
param = v
expver = 0001,
type = ef,
levelist = 1,
\end{lstlisting}
\end{multicols}

The schema defines not only the valid field identifier keys and values, but also how the FDB will internally split the identifiers provided by the user processes into three sub-identifiers which control how the Store backend lays out data in the storage system:
\begin{enumerate}
    \item \textbf{Dataset key}: describes the dataset a field belongs to. For instance, a forecast produced today starting at midday. For example, if the schema is configured to recognise \texttt{class}, \texttt{stream}, \texttt{expver}, \texttt{date} and \texttt{time} as dataset dimensions, the following dataset key would result for Listing\ \ref{listing:mars-request}:\vspace{-0.3cm}\begin{multicols}{2}
    \begin{lstlisting}[xleftmargin=-0.75cm]
    class = od,
    expver = 0001,
    time = 1200
    stream = oper,
    date = 20231201,
    \end{lstlisting}%
    \end{multicols}
    \vspace{-0.2cm}
    \item \textbf{Collocation key}: the field should be collocated in storage with other fields sharing the same collocation key. E.g.:
    \vspace{-0.3cm}\begin{multicols}{2}
    \begin{lstlisting}[xleftmargin=-0.75cm]
    type = ef,
    number = 1,
    levtype = sfc,
    levelist = 1
    \end{lstlisting}
    \end{multicols}\vspace{-0.4cm}
    \item \textbf{Element key}: identifies the field within a collocated dataset. E.g.:
    \vspace{-0.3cm}\begin{multicols}{2}
    \begin{lstlisting}[xleftmargin=-0.75cm]
    step = 1,
    param = v
    \end{lstlisting}
    \end{multicols}\vspace{-0.4cm}
\end{enumerate}

The FDB API has precisely determined semantics. Aside from a number of administrative functions, there are four primary functions in the API, namely \verb!archive()!, \verb!flush()!, \verb!retrieve()!, and \verb!list()!. The Store and Catalogue backends are responsible for ensuring that the correct semantics are provided on top of whatever provisions are made by the underlying storage systems. In particular\cite{fdb-pasc2019}:

\begin{enumerate}
    \item Data is either visible, and correctly indexed, or not. The FDB adheres to the ACID\cite{ACID} (Atomicity, Consistency, Isolation, and Durability) semantics commonly used to define database transactions.
    \item \verb!archive()! blocks until the FDB has taken control of (a copy of) the data. Data is not necessarily visible to consumers or persisted in storage devices at this point, but it is permitted to be.
    \item \verb!flush()! blocks until all data \verb!archive()!ed from the current process is persisted into the underlying storage medium, correctly indexed and made visible and accessible to any reading process via \verb!retrieve()! and \verb!list()!.
    \item Once data is made visible, it is immutable.
    \item Data can be replaced by \verb!archive()!ing a new piece of data with the same metadata. This second \verb!archive()! shares the semantics of the first, such that the old data is visible until the new data is fully persisted and indexed.
\end{enumerate}

\section{The Distributed Asynchronous Object Store (DAOS)}

The Distributed Asynchronous Object Store (DAOS)\cite{daos-scfa2022} is an open-source high-performance object store designed for massively distributed Non-Volatile Memory (NVM), including Storage Class Memory (SCM) which resides in the memory domain of a compute node. DAOS was originally developed by Intel and has recently been transferred to the DAOS foundation\cite{daos-foundation}. It provides a low-level key-value storage interface on top of which other higher-level APIs, also provided by DAOS, are built. Its features include transactional non-blocking I/O, fine-grained I/O operations with zero-copy I/O to SCM, end-to-end data integrity, and advanced data protection. The OpenFabrics Interfaces (OFI) library is used for low-latency communications over a wide range of network back-ends.

DAOS is deployable as a set of I/O processes or engines, generally one per physical socket in a server node, each managing access to NVM devices associated with the socket. Graphical examples of how engines, storage devices and network connections can be arranged in a DAOS system can be found at \cite{daos-architecture}.

An engine partitions the storage it manages into targets to optimize concurrency, each target being managed and exported by a dedicated group of threads. DAOS allows reserving space distributed across targets in \textit{pools}, a form of virtual storage. A pool can serve multiple transactional object stores called \textit{containers}, each with their own address space and transaction history.

The low-level key-value DAOS API is provided in the \texttt{libdaos} library, and it exposes what is commonly known as a map or dictionary data structure, with the feature that every value indexed in it is associated to two keys: the distribution (\textit{dkey}) and the attribute (\textit{akey}). All entries indexed under the same dkey are collocated in the same target, and the akeys identify the different entries under a same dkey. Listing dkeys and akeys is supported. This is an advanced API and not commonly used in third-party libraries using DAOS. High-level Key-Value and Array APIs are also provided as part of \texttt{libdaos}, both building on top of the low-level API.

The high-level Key-Value API exposes a single-key dictionary structure, where limited-length character strings (the \textit{keys}) can be mapped to byte strings of any length (the \textit{values}). Entries can be added or queried with the transactional \texttt{daos\_kv\_put} and \texttt{daos\_kv\_get} API calls. Querying the size of an entry and listing keys is also supported.

The Array API is intended for bulk-storage of large one-dimensio\-nal data arrays. An in-memory buffer can be stored into DAOS or populated with data retrieved from DAOS with the transactional \texttt{daos\_array\_write} and \texttt{daos\_array\_read} API calls. These operations support targeting one or multiple byte ranges with arbitrary offset and length.

Upon creation, Key-Value and Array objects are assigned a 128-bit unique object identifier (OID), of which 96 bits are user-managed. These objects can be configured for replication and striping across pool targets by specifying their \textit{object class}. If configured with striping, they are transparently stored by parts in different low-level dkeys and thus distributed across targets, enabling concurrent access analogous to Lustre file striping.

DAOS also distributes a \texttt{libdfs}\cite{libdfs-dfuse} library which implements POSIX directories, files and symbolic links on top of the described APIs, such that an application including this library can perform common file system calls which are transparently mapped to DAOS. \texttt{libdfs} is, however, not fully POSIX-compliant and cannot support the FDB POSIX Catalogue backend. A FUSE daemon and interception library are also distributed by DAOS for use in existing applications using the POSIX file system API without modification.

An important feature of DAOS, if compared to POSIX distributed file systems, is that contention between writer and reader processes is resolved server-side with a lockless mechanism rather than via distributed locking on the clients.

POSIX mandates that all write and read operations must be consistent. That is, a read operation initiated right after a write operation on the same file extent must see the data being written by the write operation. And a write operation initiated right after a read operation must not modify the original data before it is fully returned to the reader. In distributed file systems these guarantees are commonly accomplished by having a distributed locking mechanism such that every process starting a write or read operation must request a write or read lock from a lock server for the target file extent before writing or reading the extent from storage, and in case of conflicting I/O the last racing process blocks until it obtains the lock it requested. Any issued set of conflicting write and read operations is thus guaranteed to be consistent and with no interruptions or failures due to I/O conflicts. Note that every lock request involves a network round-trip to the lock server.

In DAOS, instead, a Multiversion Concurrency Control (MVCC) method is used. When a write operation is issued, it is immediately persisted by the server in a new region or object in storage, with no read-modify-write operations. The new object is then atomically indexed in a persistent index, usually in low-latency SCM distributed across the DAOS servers, and the write operation returns successfully. Any subsequent read operation for that object triggers visitation of the index and returns the associated data from the corresponding storage regions across the servers. This way, writes always occur in new regions without modifying data potentially being read, and reads always find the latest fully written version of the requested object in the corresponding index entry. This mechanism not only ensures strong consistency guarantees but also avoids the use of locks and the associated overheads.

Another relevant feature of DAOS is that metadata operations are performed against all server nodes in DAOS, rather than on dedicated metadata servers which can potentially bottleneck and hinder performance at scale. Previous work by the authors\cite{daos-ipdps} has demonstrated that DAOS provides a sufficient level of consistency and performance under contention in simplified NWP I/O workflows.

In terms of authentication and authorisation, the DAOS pools and containers can be configured via Access Control Lists (ACLs) to allow read-only or write-read access to different users. The client processes attach their effective UNIX user and group in every I/O operation sent to the server, and these are used to determine access permissions according to the ACLs.

\section{Design of the DAOS FDB Backends}

The FDB internally implements indexing functionality in what is known as a Catalogue backend, and functionality for storage (bulk write and read) of meteorological objects in a Store backend. The FDB defines abstract interfaces for these backends, such that specific Catalogue or Store instances can operate on top of a given type of storage system. If backends conform to the established interfaces and semantics, the FDB will guarantee its external API semantics. Any pair of conforming Catalogue and Store backends can then be used in conjunction even if they operate on different underlying storage systems.

Both backend interfaces define \verb!archive()!, \verb!flush()! and \texttt{retrieve()} methods, and a \verb!list()! method is also defined for the Catalogue backend interface. A call to the high-level FDB API will result in internal calls to the corresponding methods in the lower-level backend interfaces. For example, an FDB \verb!flush()! call will internally call both the Store \verb!flush()! and the Catalogue \verb!flush()!.

The DAOS Catalogue and Store backends have been implemented according to the design of the Field I/O benchmark, developed during ECMWF's preliminary assessment of DAOS\cite{daos-ipdps}. The primary idea being to use the DAOS Key-Value and Array APIs to build an indexing structure and data store, where all contention and consistency management is handled by DAOS on the server side. A graphical representation of the different DAOS entities involved is shown in Fig. \ref{fig:fdb_daos_backend}. All dataset, collocation or element keys are stringified for indexing by joining all values in the key with a ':' character, which can symmetrically be used to reconstruct the key.

\subsection{The Store Backend}

\subsubsection{Interface}

A Store backend implements an \verb!archive()! method accepting a pointer to in-memory data, a dataset key, and a collocation key. The data is taken control of and optionally persisted into storage before the method returns. A field location descriptor (equivalent to a URI) is returned describing where the data is to be persisted. This location should be collocated with other fields sharing the same collocation key. The storage location must be unique, avoiding collisions with concurrent processes. \verb!archive()! will be called repeatedly with the same dataset and collocation key, and previously archived fields must not be overwritten or modified.

The \verb!flush()! method blocks until all data which has been \texttt{ar\-chive()}ed has been persisted to permanent storage and made accessible to external reading processes.

Given a field location descriptor, the \verb!retrieve()! method builds and returns a \verb|DataHandle| (a backend-specific instance of an abstract reader object), allowing the calling process to read the field from storage without knowledge of the backend implementation.

\subsubsection{Implementation}

\begin{figure*}[htbp]
    \centering
    \includegraphics[width=490pt]{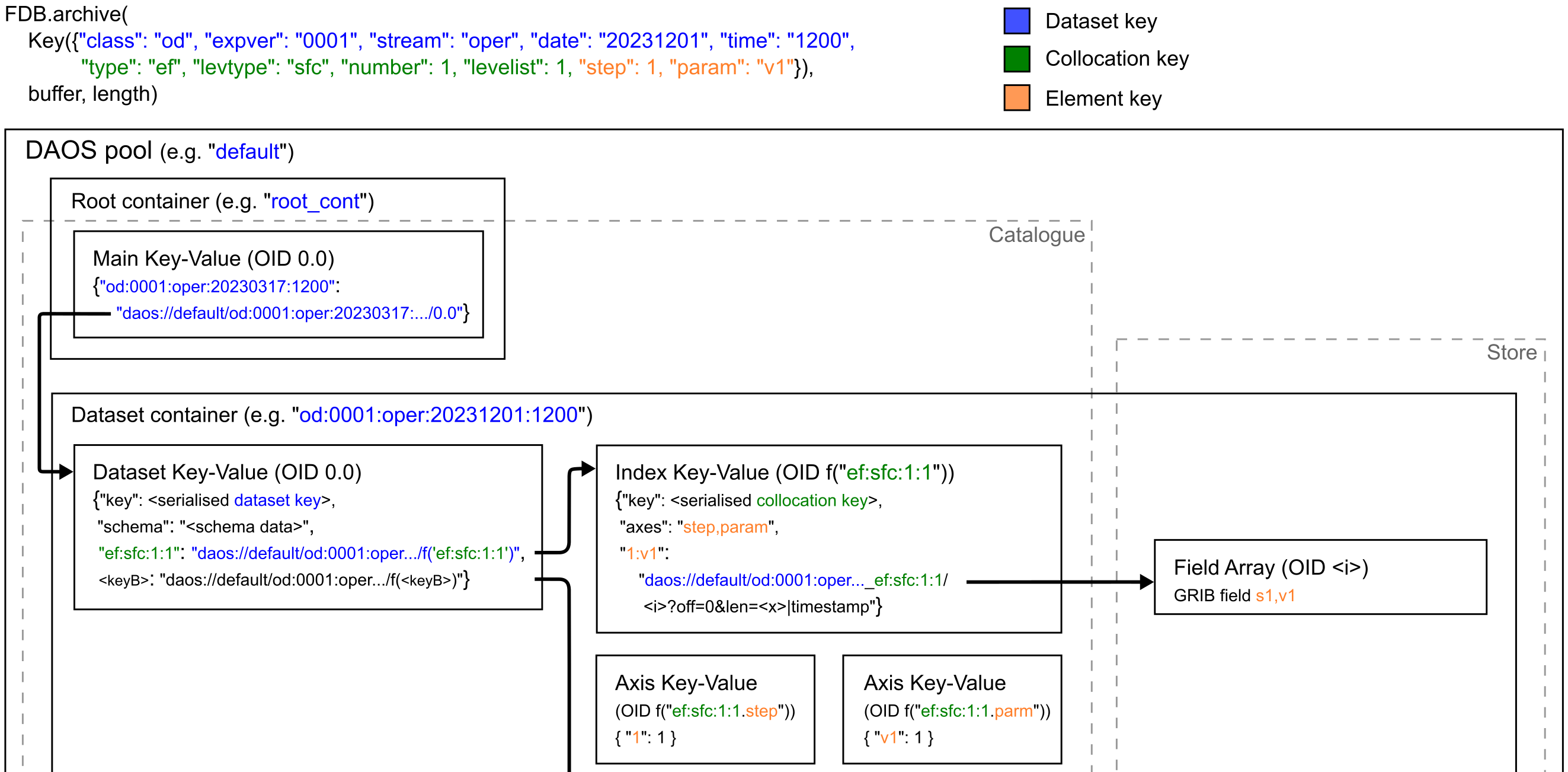}
    \caption{Diagram of DAOS entities resulting from an FDB archive call to store and index the first weather field in a simulation.}
    \label{fig:fdb_daos_backend}
\end{figure*}

Data is stored in the DAOS backend in containers identified by a stringified representation of the dataset key. When \verb|archive()| is called, the container is created if it does not exist. Once opened for use the relevant DAOS handle is cached for the process lifetime.

Every object is \verb|archive()|ed by a single process into its own DAOS Array object. The Array is created and opened with a unique object identifier (OID) obtained from DAOS. Allocating unique OIDs requires a round trip to the DAOS server to avoid collisions, and as such a range of OIDs are pre-allocated and cached by a process before object creation. Data is written into the array and the array closed. A unique field location descriptor to this array, also containing a length and an offset, is then returned. As this backend uses an Array for each object, the offset is always zero.

By contrast to the POSIX backend implementation, as the DAOS API immediately persists objects and makes them available, the DAOS Store correspondingly makes objects available to external readers immediately, and there is no further action to be taken in the \verb|flush()| method. In the future, the \verb|flush()| method may be useful if (for example) the non-blocking DAOS APIs are used in which case the flush method would block until those operations were complete.

Note that the input collocation key has not been used to determine data placement. In a previous version of the backend this key was used to create separate containers that collocated data, but the additional containers resulted in significant performance overheads so they have been removed. As such, all the object data associated with a single dataset key is collocated in the same container. The collocation key is nonetheless still used in the Catalogue backend for structuring the indexing information.

When \verb!retrieve()! is called, the input field location descriptor is used to build a \verb|DataHandle| object which reads data from the specified DAOS Array. Note that no call needs to be made to DAOS (and thus over the network to the DAOS server) to obtain the array size, as that is encoded in the field location descriptor.

\subsection{The Catalogue Backend}

\subsubsection{Interface}

A Catalogue backend has more complexity than a Store backend, not only because it addresses a more complex problem --- that of maintaining a consistent index under contention --- but also because it is involved in more operations, including archival, retrieval, removal, and listing. The Catalogue interface comprises more than ten methods, but only the ones required for field archival and retrieval are covered here, with some simplification.

Catalogue backends require an \verb!archive()! method such that, given a dataset key, a collocation key, an element key, and a field location descriptor (from an earlier Store \verb!archive()! call) the descriptor is inserted in an indexing structure, either in memory or in storage, before the method returns (with no return value). That indexing structure maps the element key to the provided associated descriptor location. This may be an internal operation, and does not guarantee persistence or visibility to reading processes. The backend may make use of the dataset and collocation keys, and knowledge of the writing processes to index related information together, or separately, as makes most sense for performance on the targeted data subsystems.

\verb!flush()! must also be implemented, which blocks until all indexed information has been persisted and made visible to any external \verb|retrieve()|ing processes. One of the requirements is that all \verb!flush()!ed indexing structures should also be accessible to \verb!list()!ing processes in a reasonably efficient manner. This may require additional indexing structures to be present such that all entries can be explored from a single entry point.

The index must \emph{always} be consistent from the perspective of an external reading process, even under read and write contention. If multiple \verb!archive()! calls with same dataset, collocation and element keys occur, the older data should be replaced by newer in a transactional manner from the perspective of any reading processes.

The \verb!retrieve()! method must return the field location descriptor given the dataset, collocation and element keys for a field. This can then be passed as an argument to the Store \verb!retrieve()! method to retrieve the actual data. It is important to note that, due to the potential for the use of the FDB as a cache in a larger data infrastructure, failing to find a field is not an error and results only in no data being returned.

The \verb!list()! method, although not covered in detail here, must return a list of field identifiers and location descriptors matching a partial request comprising spans of possible key values (i.e. not necessarily specifying all keys required for an identifier according to the schema). Making this listing possible and efficient added further constraints on the implementation approaches and choices.

\subsubsection{Implementation}

The DAOS Catalogue backend is built using a network of DAOS Key-Value objects to construct a navigable index of the stored fields. To make the data explorable, a Key-Value object identified with OID \verb|0.0| is built, as a single point of entry, in a configured root container. Below this, each dataset is assigned its own container. Inside this container a first level (dataset) Key-Value object, also with OID \verb|0.0|, maps collocation keys onto the second level (index) Key-Value object, which in turn maps element keys onto data location descriptors.
  
In parallel with this, a set of Key-Value objects (the Axis Key-Values) is built per index Key-Value, describing the span of values indexed in it. One Key-Value object is created for each element key component to act as set containing all the values written at that level. This can be used to improve the efficiency of the read process.

When \verb|archive()| is called the root pool and container are opened. If the (stringified) dataset key is not found in the root Key-Value object, the dataset container is created (with the same name as used in the Store backend), and populated with a dataset Key-Value object at OID \verb|0.0|. An entry is placed in the root Key-Value mapping the dataset key to the (URI of the) dataset Key-Value. As there is a meaningful overhead to handling pools and containers, once any have been opened they are cached for the lifetime of the process.

The collocation key is handled in the same manner. If the (stringified) collocation key is not found in the dataset Key-Value object a new Key-Value is created identified by the collocation key, and an entry added to the dataset Key-Value mapping the collocation key to the URI of the index Key-Value. Otherwise, the identified Key-Value is used. An entry is then added to index Key-Value mapping the element key to the supplied field location description.

After \verb!archive()! has returned, the indexing information has already been persisted and the indexed fields have been made visible to reader processes. There is no further action to be taken in the \verb!flush()! method.

Concurrent processes writing or reading fields with same dataset and collocation keys will contend on a same index Key-Value, which can have a noticeable performance overhead. It is therefore advisable to configure the FDB schema such that as few parallel processes as possible share the same keys, as is done in the NWP I/O workflows considered in this paper. Alternatively, it would be possible to extend this design such that each writing process uses its own unique Key-Value, in a similar way to the POSIX catalogue backend, although this would significantly increase implementation complexity.

The root and dataset Key-Values in this implementation primarily make the fields efficiently findable and listable. By having its own container, entire datasets can be easily and efficiently removed, facilitating the use of the FDB as a rolling archive.

When \verb!retrieve()! is called, the root pool and container are opened. The dataset key is looked up in the root Key-Value to identify the corresponding container and dataset Key-Value. The collocation key is then looked up in this dataset Key-Value to identify the index Key-Value in the same way. Finally the element key is looked up in the index Key-Value and, if found, the field location descriptor is returned.

The \verb!list()! method makes use of the information stored in the Axis Key-Values. The Axis object is compared against the query and the index Key-Value is skipped if nothing contained in it matches the query. As axis information is cached in the calling process, it may become out of date and need to be purged, but the FDB guarantees that any fields written before the last \verb|flush()| will be correctly visible to any tasks started after.

The transactionality of the \texttt{daos\_kv\_put} and \texttt{daos\_kv\_get} operations on the Key-Value objects is critical to ensuring consistency of indexes under \verb!archive()! and \verb!retrieve()! contention, and to ensuring that such contention is resolved on the DAOS server.

Although any concurrent writer and reader processes interact with the root and dataset Key-Values, contention on these Key-Values is avoided by caching the relevant entries in the reader process pathway, and excessive load from \verb!archive()!ing processes can be avoided by using only a few different dataset and collocation keys in every writer's lifetime. These approaches and design decisions mean that index Key-Values remain as the objects where most of the contention occurs.

\section{Performance assessment}

Performance and scalability tests have been carried out comparing the new DAOS FDB backend to the existing POSIX backend using DAOS and Lustre installed on the same hardware resources. The same test configurations were also run through the Field I/O benchmark\cite{daos-ipdps}, a standalone benchmark tool developed to evaluate the performance of DAOS without involving the full complexity of ECMWF's I/O stack while reproducing the I/O operations and patterns the DAOS backend requires.

\subsection{Test system}

Tests have been carried out on NEXTGenIO\cite{ngio}, a research HPC system composed of 34 dual-socket nodes with Intel Xeon Cascade Lake processors. Each socket has six 256 GiB first-generation Intel\textsuperscript{\textregistered} Optane\textsuperscript{\texttrademark{}} Data Centre Persistent Memory Modules\cite{intel-3dxpoint}\cite{hmem-3dxpoint} (DCPMMs) configured in AppDirect interleaved mode. There are no NVMe devices. Each processor is connected via its own integrated network adapter to a low-latency OmniPath fabric. Each of these adapters has a maximum bandwidth of 12.5 GiB/s.

The fabric is configured in dual-rail mode. That is, a separate network for each processor socket on a node, meaning there are two high performance networks per node. The HPC system nodes use CentOS7 as the operating system, with DAOS v2.4.

For the DAOS benchmarking, DAOS has been deployed on different numbers of storage nodes with a single DAOS engine per socket (i.e. two per node), using the full ext4 file system on the Optane SCM for that socket. Each socket uses the associated fabric interface and interleaved SCM devices, with 12 targets per engine. DAOS does not support using PSM2, a low-latency communication protocol implementing RDMA on OmniPath. Instead, it was configured to use the TCP protocol.

Lustre has been deployed as well on different numbers of storage nodes, with one OST per socket, plus one node devoted to the metadata service. Both the OSTs and the MDTs used in the file system mount an ext4 file system on the SCM attached to their respective sockets, providing 1.5 TB of high-performance storage per OST and MDT, with servers and clients connected using the high performance PSM2 communication library.

Up to 20 nodes have been employed to execute the benchmark client processes using both sockets and network interfaces. SCM in the client nodes was not used and did not have any effect on I/O performance.

\subsection{Benchmark}

\verb!fdb-hammer! is an FDB performance benchmarking tool provided in the FDB Git repository\cite{fdb-github}, which can be built alongside the other FDB command-line tools.

\verb|fdb-hammer| takes a single GRIB field, supplied on the command line, and uses its encoded metadata as a template to generate a sequence of fields to be archived, retrieved or listed. \verb|fdb-hammer| processes are independent without synchronisation with any other parallel processes. In this manner it simulates the behaviour of model I/O server processes which write independent streams of data to the FDB, and the behaviour of post-processing tools which also read independently. It creates an ``I/O pessimised'' benchmark --- i.e. a worst possible case for I/O as all relevant computation has been removed.

\verb|fdb-hammer| is configured with a series of command line arguments (including \verb|--class|, \verb|--expver| and \verb|--nlevels|) specifying the span of metadata to be iterated during the run. Most significantly \verb|--nsteps| controls the values of the \verb|step| keyword in the metadata and therefore, if \verb!fdb-hammer! is invoked in its archive mode, it also controls the number of times \verb|flush()| is called and the number of fields written in sequence between flushes.

A number of benchmark scripts\cite{field-io} have been used which invoke \verb|fdb-hammer| in parallel at different scales to mimic I/O patterns in ECMWF's operational workflow. Each of the invoked \verb|fdb-hammer| processes outputs timestamps and profiling statistics which are aggregated to produce the following results.

\subsection{Methodology}

The methodology in this performance analysis is based on ECMWF's DAOS preliminary assessment with Field I/O\cite{daos-ipdps}. It considers:
\begin{enumerate}
    \item How bandwidths should be calculated
    \item Different I/O patterns of interest
    \item How to optimise benchmark and system parameters for performance and reproducibility
    \item How to scale the storage system and benchmark runs
\end{enumerate}

Regarding (1), bandwidth calculations are made from the total volume of data transferred (written or read) and the time elapsed between the first benchmark I/O start time to the last benchmark I/O end time. This is known as \textit{global timing bandwidth}.

Regarding (2), only \textit{access pattern A} (as defined in \cite{daos-ipdps}) has been considered. A writing phase is run first, where \verb!fdb-hammer! is run in parallel from a number of parallel processes in its archive mode, and each process \verb!archive()!s a sequence of fields with different metadata identifiers. Following this, \verb!fdb-hammer! is executed in \verb|retrieve()| mode from the same number of processes. There is no contention between writers and readers as they run sequentially, and so it is referred to as \textit{no w+r contention} hereafter.

A variant of that access pattern has also been considered, in which a writer phase is run initially to populate the storage, and then a writer and a reader phase are run simultaneously. In this variant there is contention between writers and readers but, in contrast to the \textit{pattern B} defined in \cite{daos-ipdps}, the processes write and read a sequence of fields with different metadata rather than repeating the same metadata. This pattern is referred to as \textit{w+r contention}.

Considering (3) Optimising the benchmark and system parameters, given such a wide possible parameter space, has been challenging. The parameter optimisation strategy outlined in \cite{daos-ipdps} has been followed in this analysis. Firstly, an appropriate run length is determined. The benchmark is run against a fixed system with fixed configuration, only varying the number of fields to be written per process, which is increased progressively until the variability in bandwidth measurements becomes small (less than 5\%). This helps balance test reproducibility whilst also keeping runtimes as short as possible.

After this, the client-to-server node ratio and the number of benchmark processes per client node are varied to find the point of diminishing returns, where the server connections approach saturation and adding additional client nodes or processes has little benefit. Note that each additional process added increases the overall I/O size of the benchmark as the fields written per process is now fixed.

Once these parameters are fixed, the I/O startup timestamps are the analysed to ensure that all processes in the benchmark are really working in parallel. If this is not true, the startup processes of the benchmark should be modified, the number of fields written needs to be increased further to minimise the impact of startup time, or explicit I/O startup synchronisation needs to be implemented.

Following this, the backend specific parameters can be tested and optimised, including in this case DAOS object class, Lustre file striping, FDB schema optimisations, and others.

Regarding (4), the number of nodes employed for the storage system can be increased progressively, and the benchmark can be run at each scale with a corresponding number of client nodes (according to the selected client-to-server-node ratio), with all access patterns of interest. The bandwidths obtained for the sequence of steps will provide insight into the scaling behaviour for the different access patterns, as more resources are employed for the storage system and the benchmark application.

\section{Results}
\subsection{Parameter Optimisation}

Following the described methodology, \verb!fdb-hammer! was run first with the DAOS backends with \emph{no w+r contention}, from 16 NEXTGenIO nodes, each node executing 32 parallel \verb!fdb-hammer! processes, and with a DAOS system deployed on 8 nodes. The size of the fields written and read was set to 1 MiB.

Writing 2000 fields per process was found to reduce variance in bandwidth measurements down to approximately 5\%. Each \verb!fdb-hammer! process was configured to make these 2000 iterations correspond to 10 steps, 10 parameters, and 20 model levels for a single dataset and member. 2000 fields per process was also found to be suitable with the POSIX backend running on Lustre deployed on 8+1 nodes. Despite this, occasional outlier benchmark runs were observed varying by more than 10\% from the mean when using the POSIX backend. To identify any potential further outliers, all tests in this paper were repeated 3 times. In the end outliers were found to be very rare and were not analysed further, and the 3 bandwidth results for each test were averaged.

\begin{figure}[htbp]
    \begin{subfigure}[b]{119pt}
        \includegraphics[width=119pt]{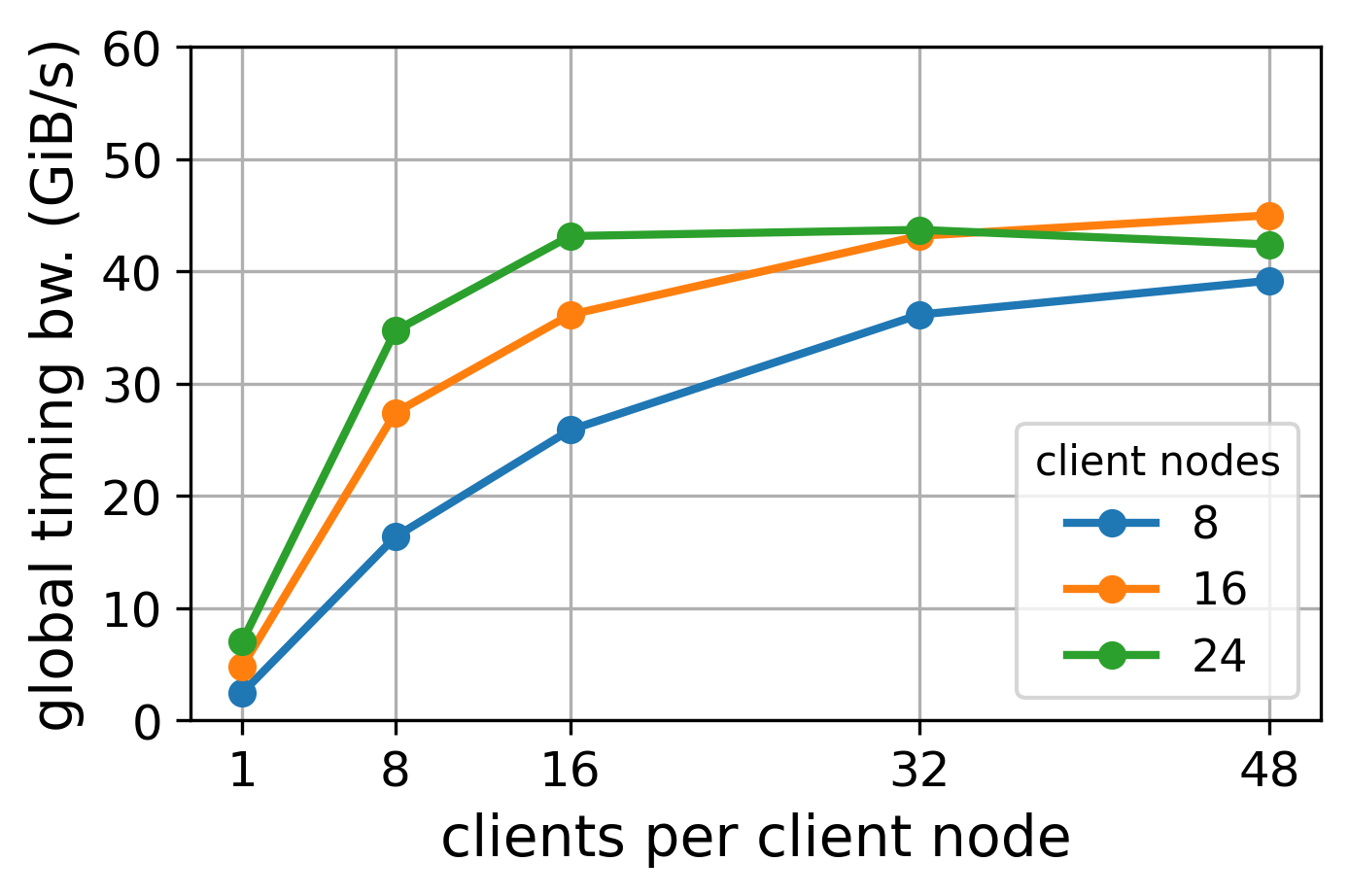}
        \caption{DAOS Writers}
    \end{subfigure}
    \begin{subfigure}[b]{119pt}
        \includegraphics[width=119pt]{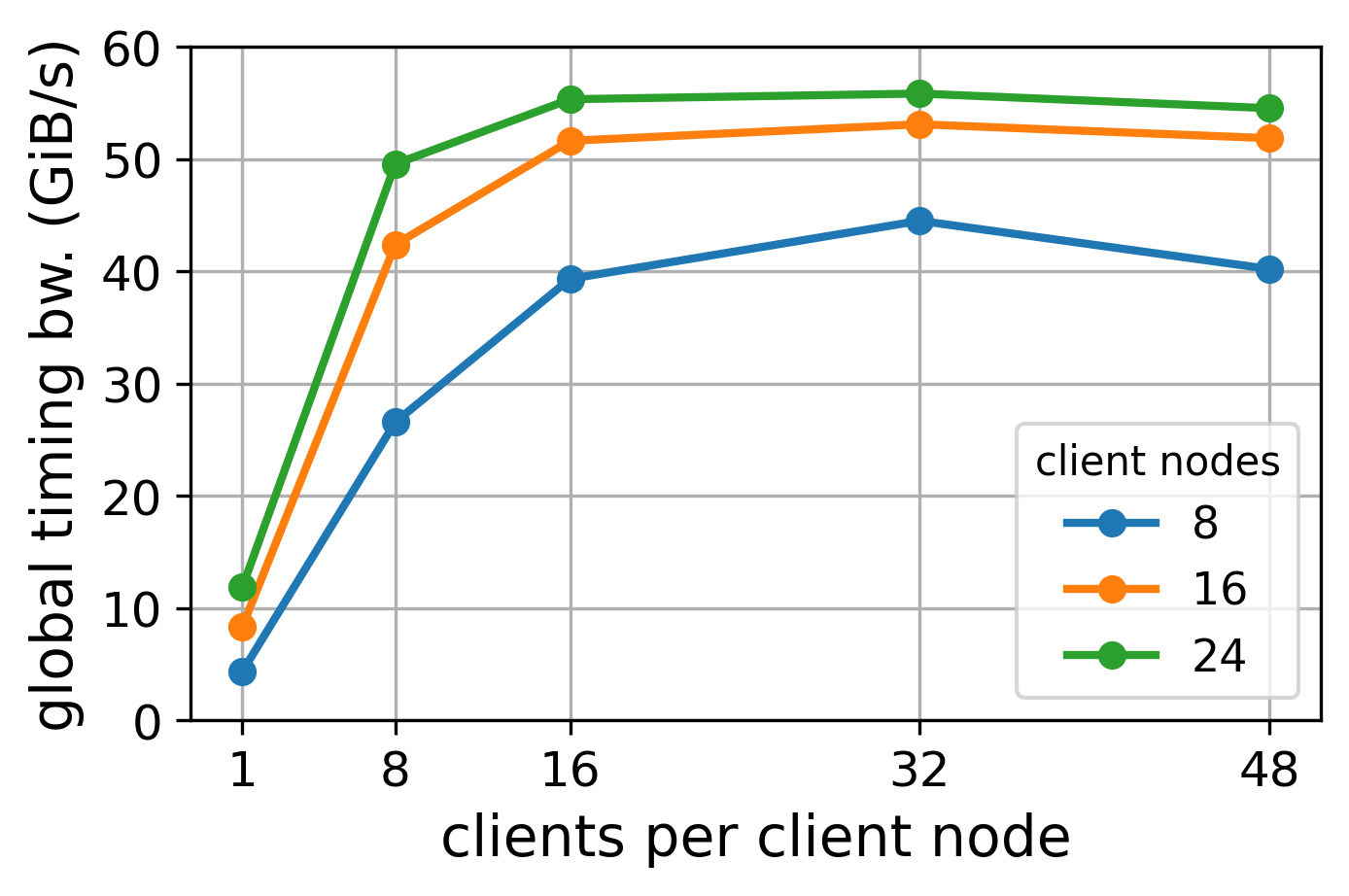}
        \caption{DAOS Readers}
    \end{subfigure}
    \vskip\baselineskip
    \begin{subfigure}[b]{119pt}
        \includegraphics[width=119pt]{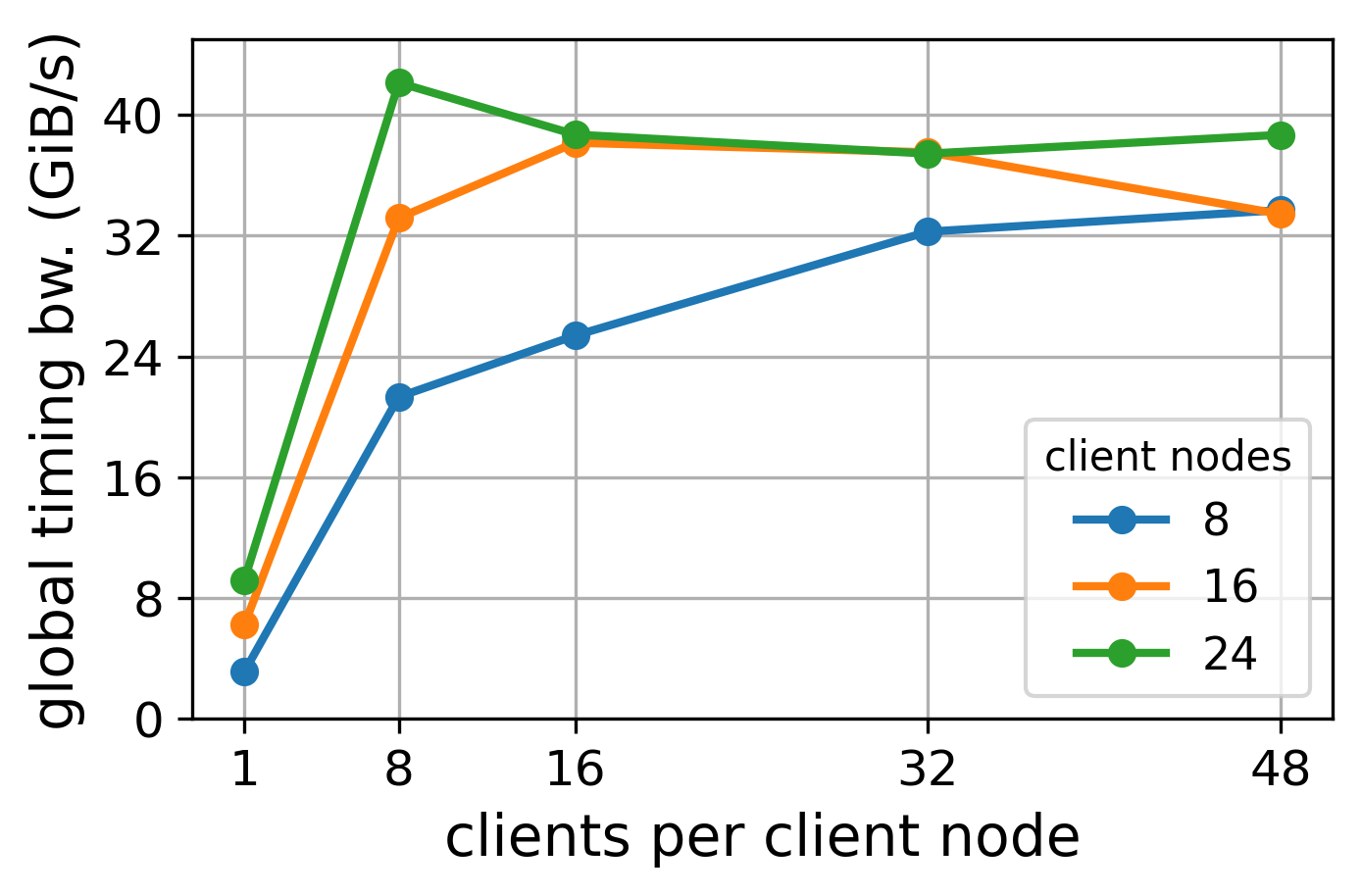}
        \caption{Lustre Writers}
    \end{subfigure}
    \begin{subfigure}[b]{119pt}
        \includegraphics[width=119pt]{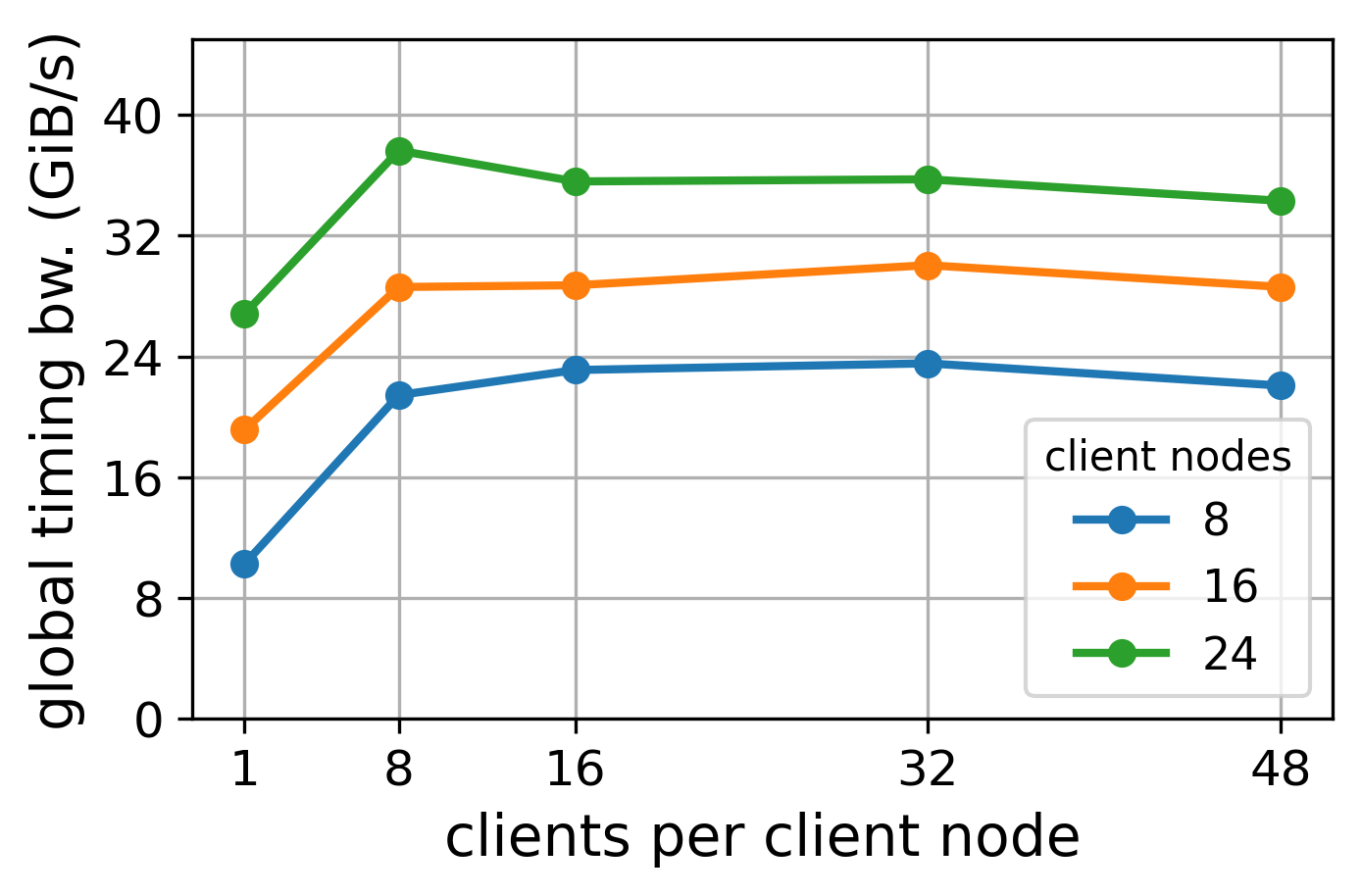}
        \caption{Lustre Readers}
    \end{subfigure}
    \caption{Bandwidth results for fdb-hammer/DAOS (8 server nodes; top row) and fdb-hammer/Lustre (8+1 server nodes; bottom row) runs. No w+r contention.}
    \label{fig:daos_lustre_cn_cpcn}
\end{figure}

Once the tests were fixed at 2000 fields per process, they were repeated, varying the client-to-server-node ratio and the number of benchmark processes per client node. The results for both the DAOS and the POSIX backends are shown in Fig. \ref{fig:daos_lustre_cn_cpcn}.

\begin{figure*}[hpb]
    \centering
    \begin{subfigure}[b]{125pt}
        \includegraphics[width=125pt]{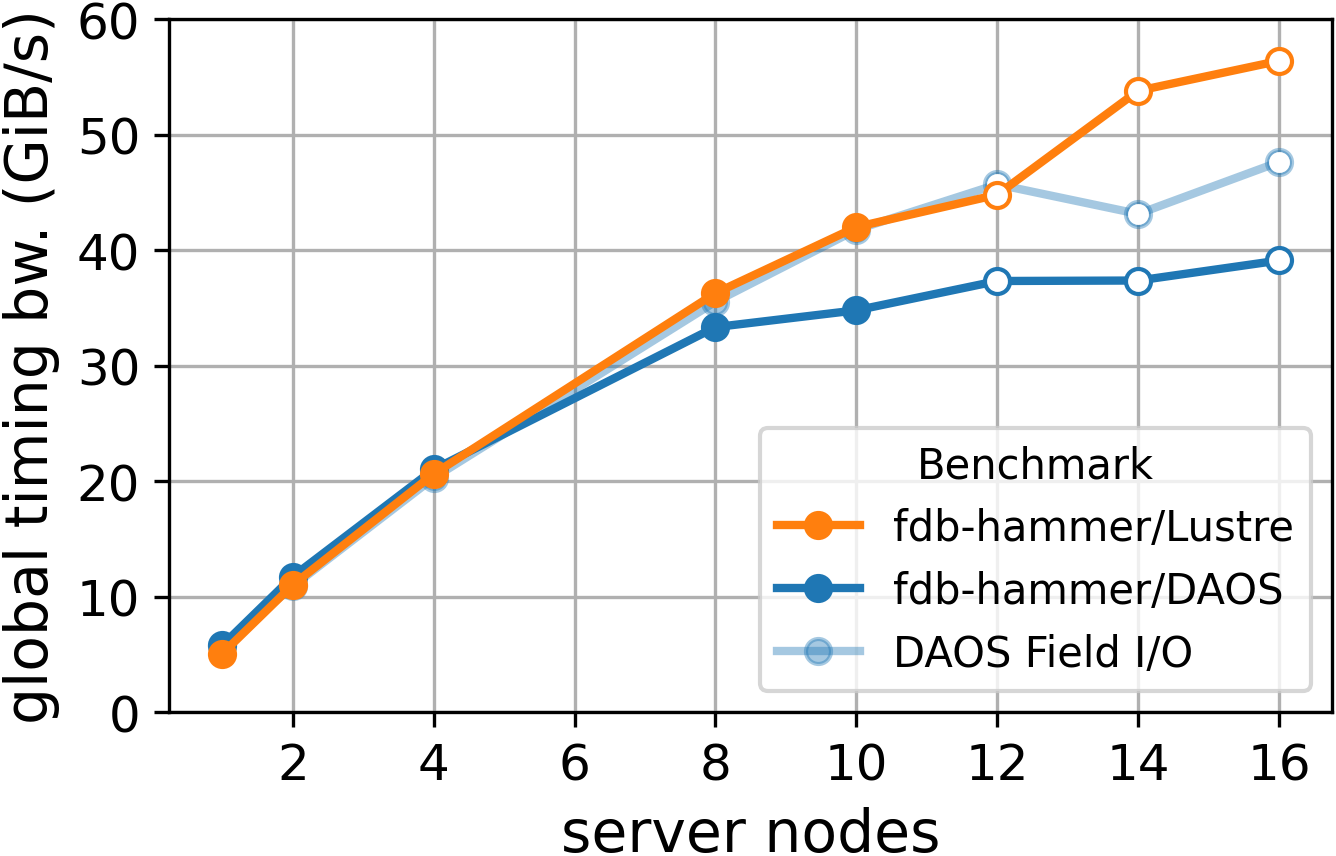}
        \caption{Writers, no w+r contention}
    \end{subfigure}
    \begin{subfigure}[b]{125pt}
        \includegraphics[width=125pt]{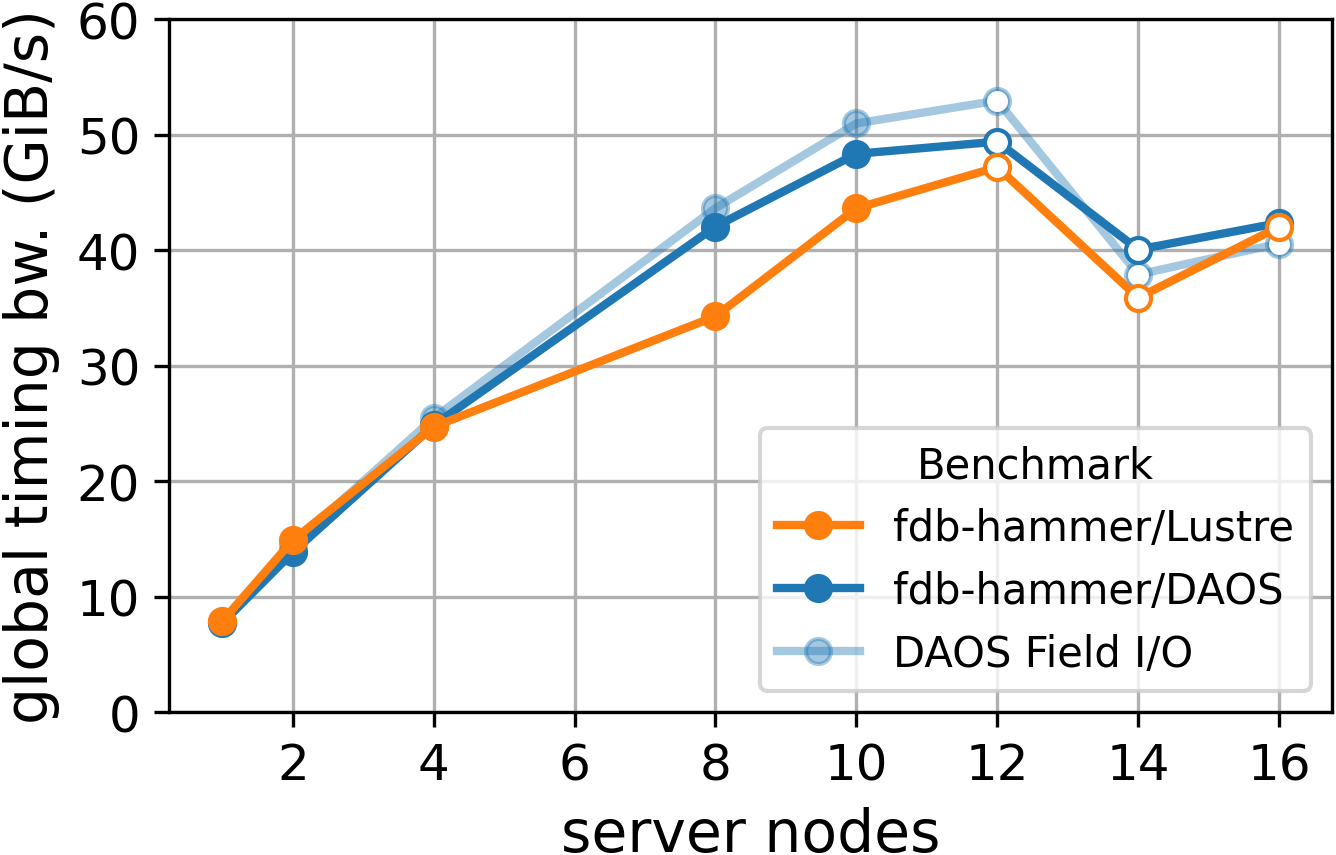}
        \caption{Readers, no w+r contention}
    \end{subfigure}
    %\vskip\baselineskip
    \begin{subfigure}[b]{125pt}
        \includegraphics[width=125pt]{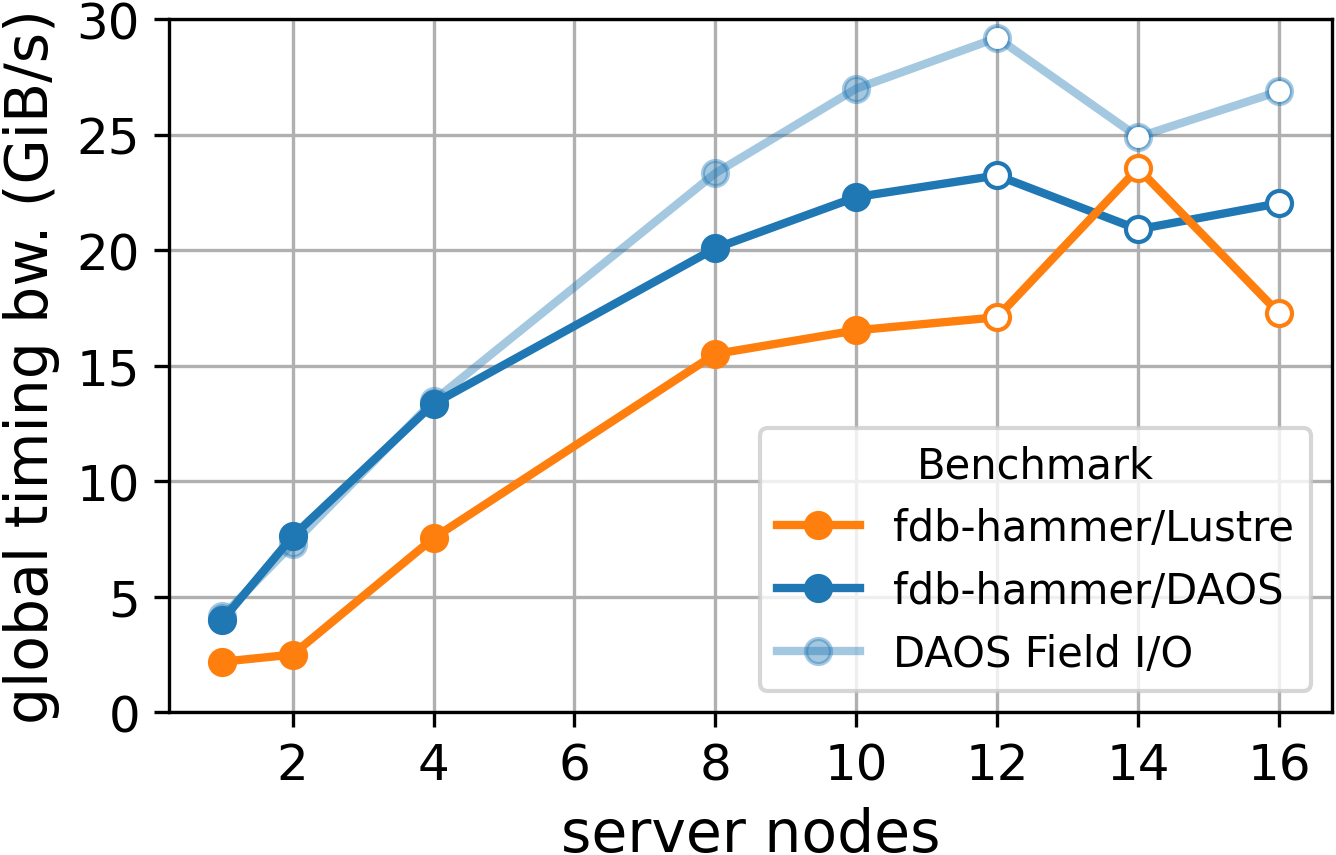}
        \caption{Writers, w+r contention}
    \end{subfigure}
    \begin{subfigure}[b]{125pt}
        \includegraphics[width=125pt]{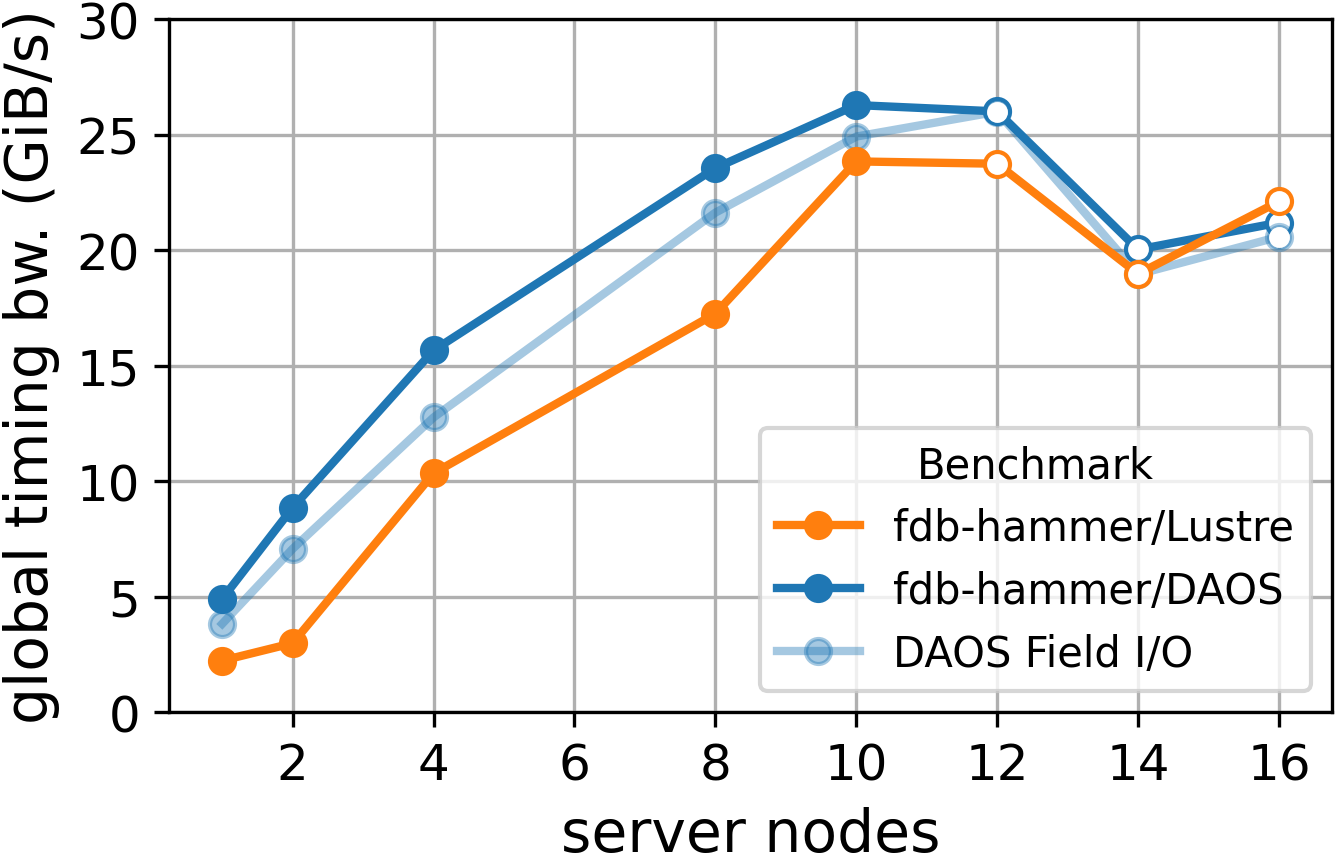}
        \caption{Readers, w+r contention}
    \end{subfigure}
    \caption{Bandwidths observed in short fdb-hammer and Field I/O runs without and with write/read contention.}
    \label{fig:scalability_short_runs}
\end{figure*}
For both DAOS and Lustre runs, a client-to-server-node ratio of 3 (i.e. 24 client nodes for the 8-server-node setup) does not result in significantly higher bandwidths compared to a ratio of 2, whereas a ratio of 2 results in substantially higher bandwidths than those obtained with a ratio of 1. This suggests a ratio of 1 is not sufficient to saturate the server-side network interface cards. A ratio of 2 has been chosen for most of the tests in the rest of the analysis.

Regarding client process counts, bandwidths generally saturate at 16 or 32 processes per node. Lustre bandwidths saturate at lower client process counts, probably because Lustre is using the PSM2 protocol --- which implements Remote Direct Memory Access (RDMA) over the OmniPath network adapters --- for efficient communications, whereas DAOS uses the TCP protocol. This issue is analysed in more detail in \cite{daos-ipdps}.

Where 8 or more server nodes have been employed, process counts of 16 and 32 have been used for all DAOS and Lustre tests. Where fewer than 8 server nodes were employed, process counts of 32 and 48 were chosen as they were found to result in the highest bandwidths.

The tests were then re-run at the largest scale allowed by the test system (20 client nodes and 12 server nodes, 12+1 for Lustre), and the timestamps output by the benchmark were analysed to identify any potential delays in I/O start up across processes. The detail of the methodology followed to address this can also be found in \cite{daos-ipdps}. No substantial delays were identified.

At this point, other relevant configuration options were tested using the \emph{no w+r contention} pattern with 20 client nodes and 12 server nodes (12+1 for Lustre). A DAOS object class of \verb!OC_S1! for DAOS Arrays resulted in the best performance, as anticipated in the preliminary assessment\cite{daos-ipdps}, likely due to the relatively small field size in conjunction with the large number of processes accessing them concurrently in a benchmark run. For the FDB schema, it was found to be optimal to have the \verb!number! and \verb!levelist! identifiers at the collocation level for the DAOS backend, as shown in Fig. \ref{fig:fdb_daos_backend}. This results in each process writing or reading from an exclusive set of index Key-Values, minimising contention. Having these dimensions at the element level performed best for the POSIX backend as writing processes already maintain independent indexes there.

It was not as straightforward to apply this optimisation methodology as this report might suggest. The DAOS backends were improved a few times, and configuration flaws were discovered and fixed both for the server and the benchmark, resulting in the procedure being repeated multiple times. Some examples of improvements and flaws that led to starting over were optimisations in the DAOS backends to avoid Array size checks in the read pathway; use of Arrays without attributes for more efficient opening and creation in the write pathway; increasing the configured number of OIDs allocated per \verb!daos_cont_alloc_oids! call in the DAOS backends; properly pinning client processes to both sockets to fully exploit the network interface cards; and using different releases of DAOS with varying performance.

\subsection{Short Scaling Tests}

Tests were run both with and without write/read contention to characterise the performance scalability of the system. The number of DAOS and Lustre server nodes employed for the \verb!fdb-hammer! runs were increased gradually from 1 to 16 (plus one in Lustre deployments for the metadata server), preserving a client-to-server-node ratio of 2 where possible. The results are shown in Fig. \ref{fig:scalability_short_runs}, with DAOS bandwidths in dark blue, Lustre bandwidths in orange, and results obtained with the Field I/O benchmark\cite{daos-ipdps} in light blue which provide insight on the bandwidths that should be achievable with the DAOS backends.

For configurations with 10 or fewer server nodes the client-to-server-node ratio of 2 was satisfied. With 12 server nodes only 20 client nodes were available, and above this a client-to-server-node ratio of 1 was used due to the limited nodes available in the system. These reduced configurations, which obtain a lower bandwidth than the server nodes could provide, have been marked in the figure with hollow dots. For the \emph{w+r contention} runs, half of the client nodes were employed as writers and the other half as readers

The DAOS and the Lustre backends perform fairly similarly, with DAOS generally achieving slightly higher bandwidths except when writing in the absence of any contention where Lustre performs best. It is encouraging that the DAOS backends attain similar bandwidths to those observed in Field I/O runs.

Results for both backends scale well but sub-linearly as the number of server nodes increases. Although the sharp decline in bandwidth at higher server node counts was expected (due to the reduced client node counts) this does not fully explain the gentle plateauing which was also observed in the Field I/O results. That plateauing was unexpected because nearly identical runs had been performed a few months ago (those reported in \cite{daos-ipdps}, Fig. 7), and no performance decline was observed then. This change in scaling behaviour in Field I/O runs is most likely due to a DAOS version upgrade from v2.0.1 in \cite{daos-ipdps} to v2.4 in this paper. We can also see from the results that DAOS provides improved performance compared to Lustre in the scenario with contention, especially for writer tasks, except where the ideal client-to-server node ratio was not fulfilled.

\begin{figure}[htbp]
    \begin{subfigure}[b]{119pt}
        \includegraphics[width=119pt]{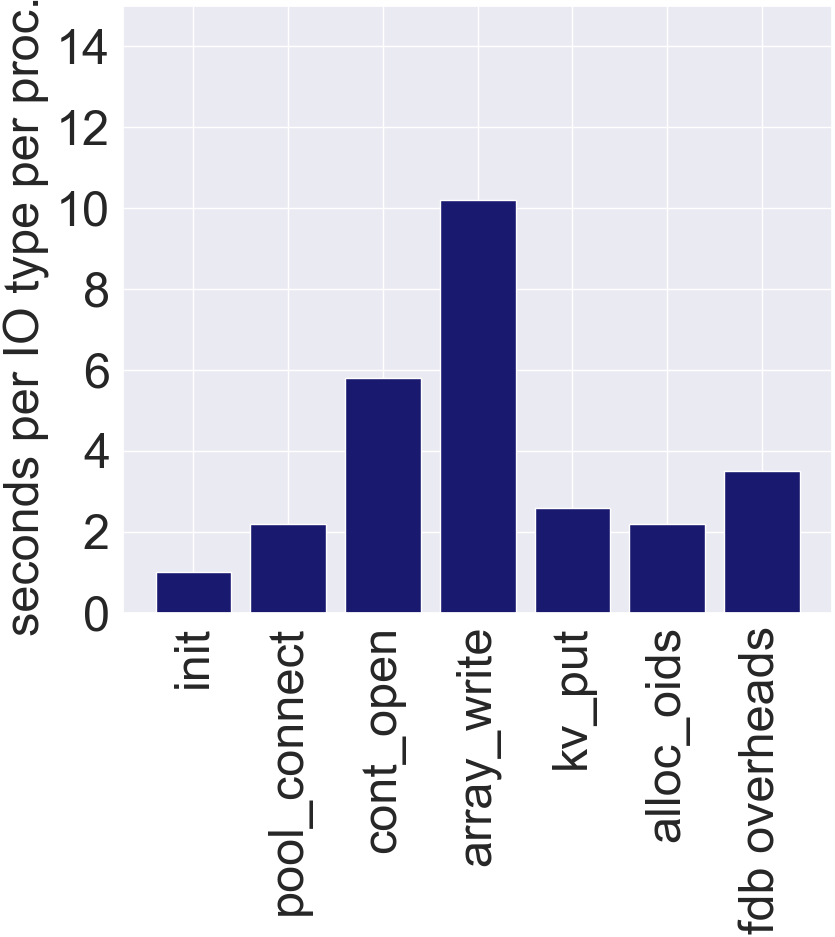}
        \caption{Writers. Average per-process wall-clock time: 27.5s.}
    \end{subfigure}
    \begin{subfigure}[b]{119pt}
        \includegraphics[width=119pt]{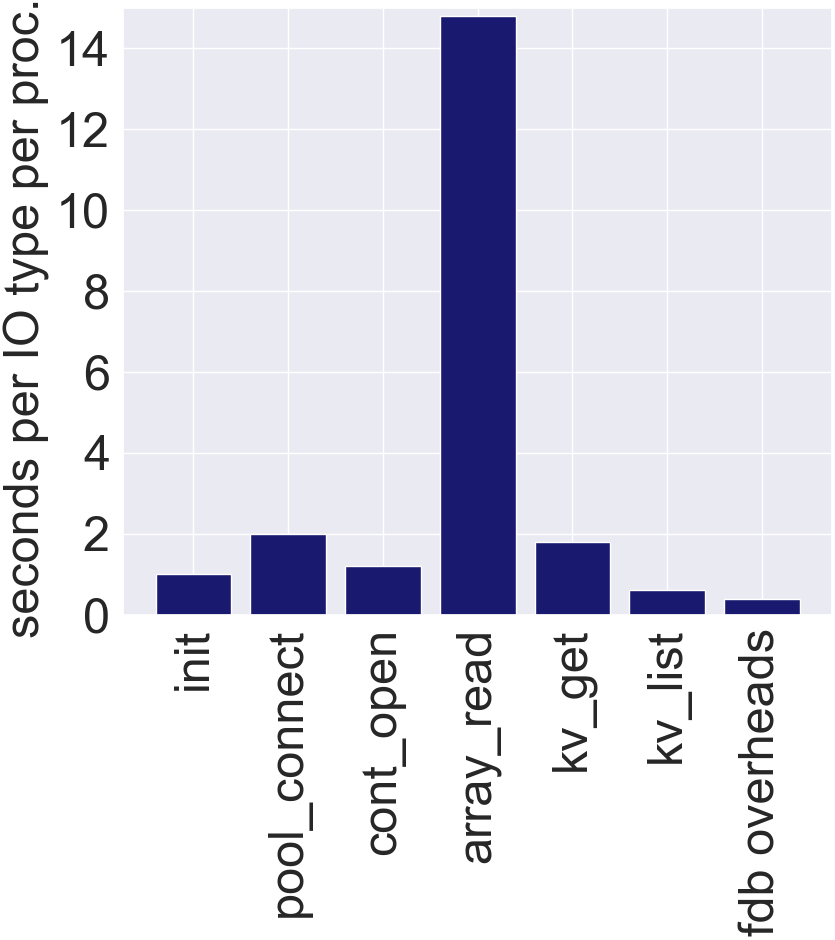}
        \caption{Readers. Average per-process wall-clock time: 21.8s.}
    \end{subfigure}
    \caption{Profiling results for fdb-hammer/DAOS runs (12 server and 20 client nodes; 32 processes per client node).}
    \label{fig:daos_bottlenecks}
\end{figure}

To get further insight into the performance characteristics of the DAOS backend, profiling information was collected from \verb!fdb-hammer! for some of the runs with 12 server nodes, which is analysed and summarised in Fig. \ref{fig:daos_bottlenecks}. Whereas most of the time was spent in \texttt{daos\_array\_write} and \texttt{daos\_array\_read} calls, which is a good sign that the transfer of data across the network and persisting into storage devices predominated, a significant amount of time is spent in one-off establishment of DAOS pool and container connections, as well as in other pool operations and other one-off FDB overheads, particularly in the \verb!fdb-hammer! writers. These overheads should become less significant in operational workloads where the processes operate for time periods one or two orders of magnitude longer than in the short \verb!fdb-hammer! runs tested.

\subsection{Longer Scaling Tests}

\begin{figure*}[htbp]
    \centering
    \begin{subfigure}[b]{125pt}
        \includegraphics[width=125pt]{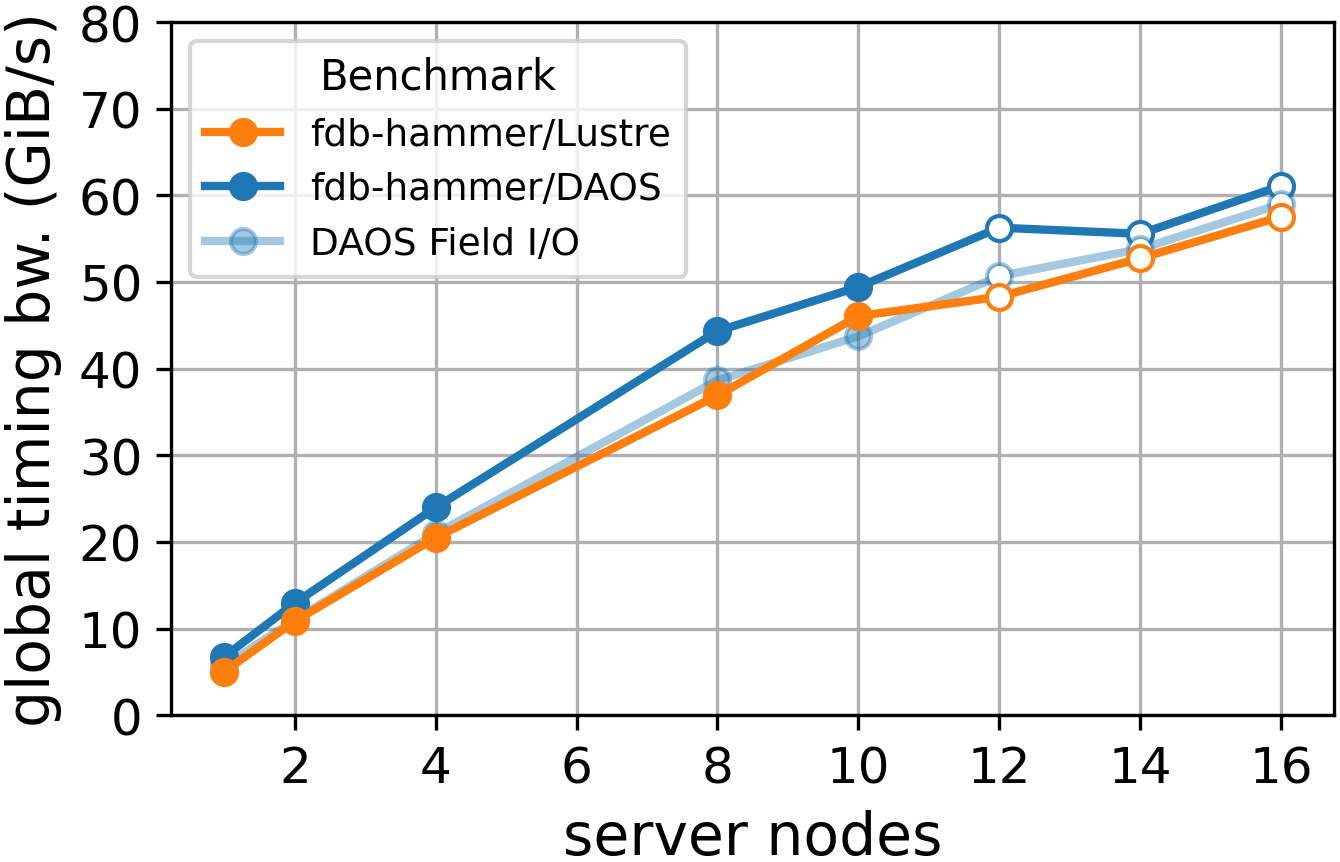}
        \caption{Writers, no w+r contention}
    \end{subfigure}
    \begin{subfigure}[b]{125pt}
        \includegraphics[width=125pt]{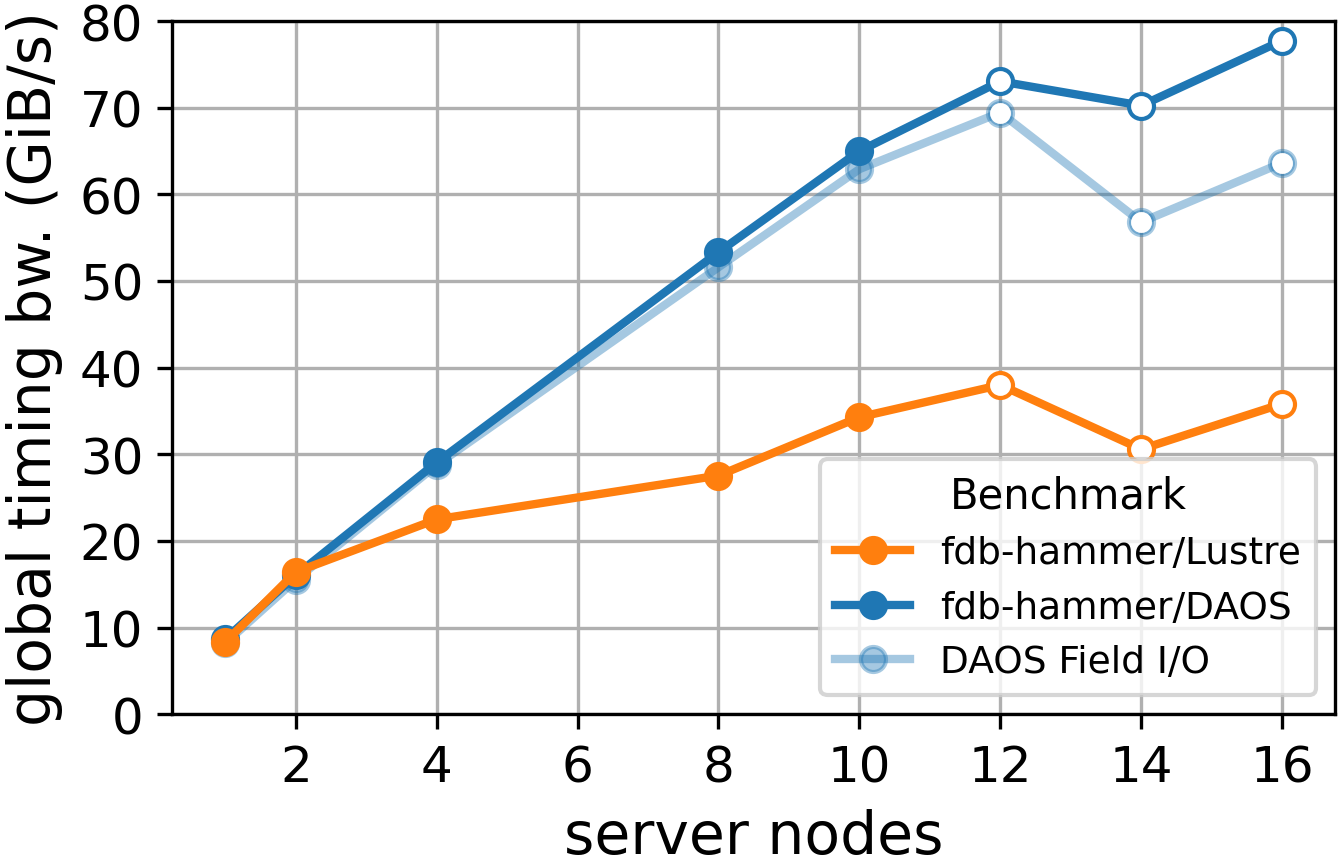}
        \caption{Readers, no w+r contention}
    \end{subfigure}
    \begin{subfigure}[b]{125pt}
        \includegraphics[width=125pt]{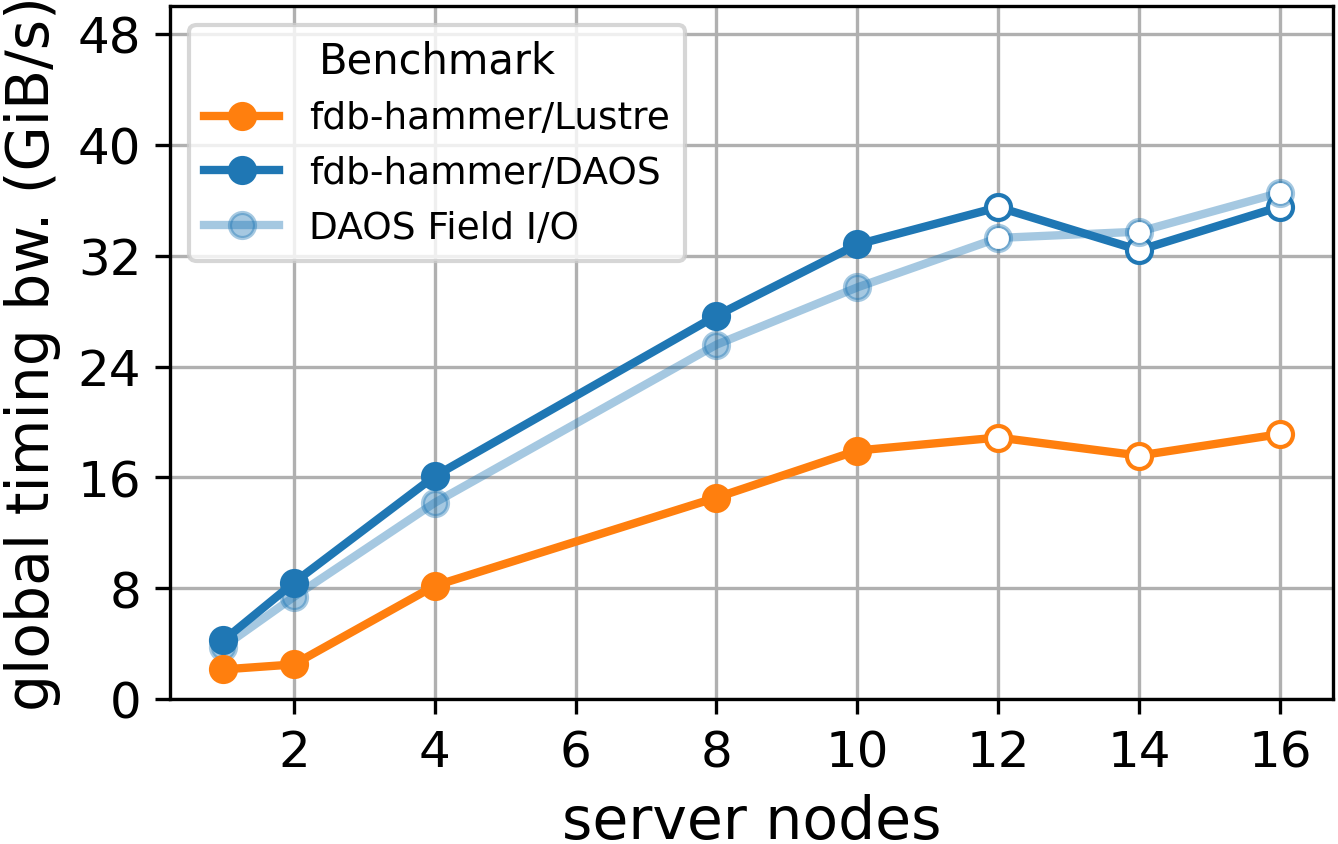}
        \caption{Writers, w+r contention}
    \end{subfigure}
    \begin{subfigure}[b]{125pt}
        \includegraphics[width=125pt]{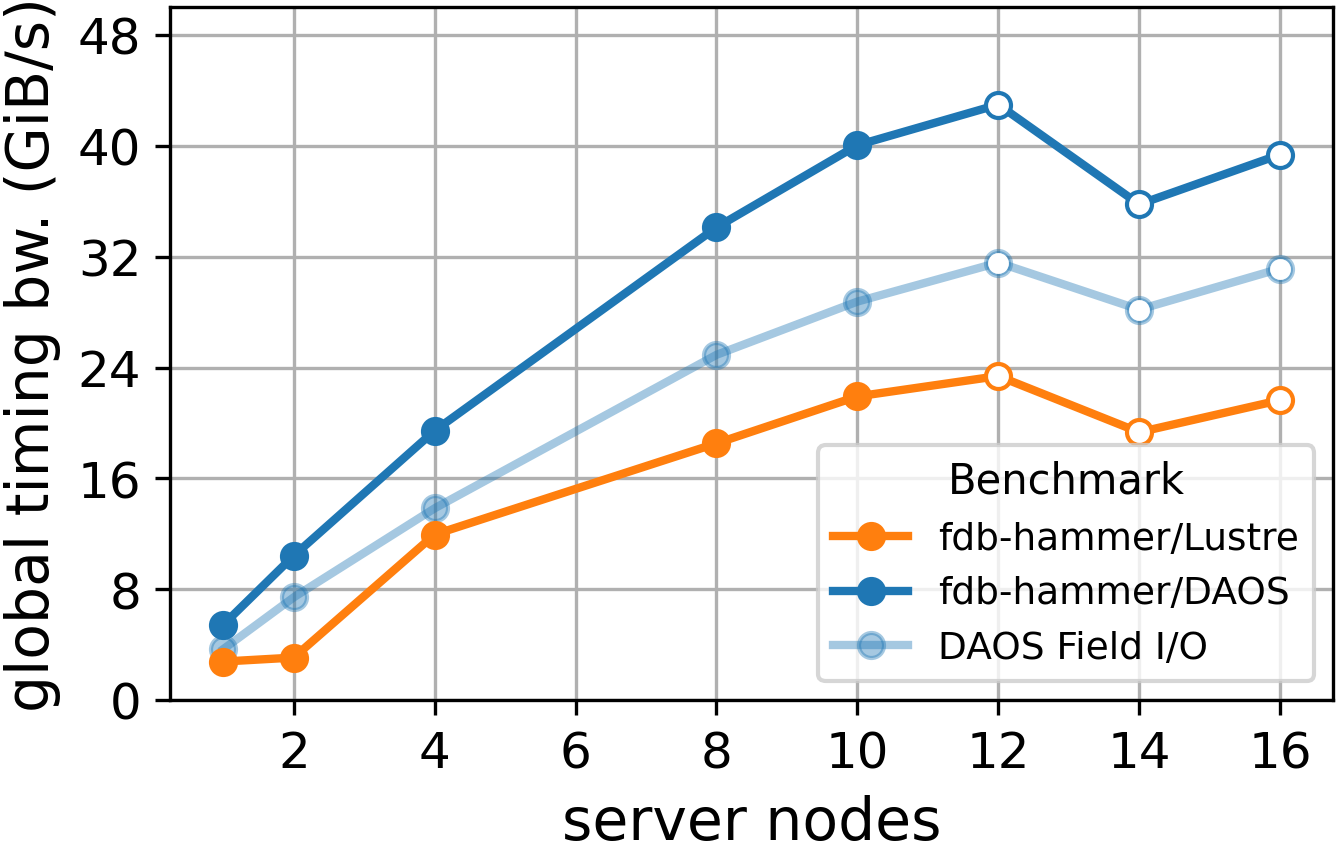}
        \caption{Readers, w+r contention}
    \end{subfigure}
    \caption{Bandwidths observed in long fdb-hammer and Field I/O runs without and with write/read contention.}
    \label{fig:scalability_long_runs}
\end{figure*}

The scalability assessment was repeated using longer runs of 10,000 fields per process, to attempt to minimise the impact of the one-off overheads and analyse something more reflective of a real I/O workload. Each \verb!fdb-hammer! process was configured to use 100 steps, 10 parameters, and 10 model levels for a single dataset and member. The results are shown in Fig. \ref{fig:scalability_long_runs}.

The longer runs scale essentially linearly in most configurations as more server nodes are added. It is interesting to see how all three benchmarks perform very similarly for (a), write with \emph{no w+r contention}, achieving similar bandwidths to the IOR/DAOS benchmarking performed on the same system and reported in \cite{daos-ipdps}. This indicates that all three benchmarks are making efficient use of the underlying storage in the absence of contention.

For (b), read with \emph{no w+r contention}, the DAOS backend again shows very good scalability reaching close-to-IOR bandwidths. The POSIX backend performs significantly worse, and this is due to the design choices made in that backend to benefit write performance at the expense of read performance, as explained in Section\ \ref{sect:fdb}.

\verb!fdb-hammer! with the DAOS backends generally reaches slightly higher bandwidths than were achieved with Field I/O. This is due to some optimisations implemented in the FDB DAOS backends which were not implemented in Field I/O, particularly the use of \texttt{daos\_array\_open\_with\_attrs} on write and avoiding unnecessary \texttt{daos\_array\_get\_size} calls on read.

In (c) and (d), with \emph{w+r contention}, DAOS performs remarkably well with nearly linear scaling, whereas Lustre shows 50\% lower bandwidths with a marked performance decline starting at 4 server nodes, presumably due to file locking overheads.

These excellent results demonstrate that the new object store semantics and contention resolution mechanisms in DAOS can enable high I/O performance at scale in ECMWF's operations and potentially in other applications with strong I/O contention.

Although the full contention and the reader view transposition in operational NWP, described in Section\ \ref{sect:nwp}, could not be reproduced exactly in the benchmarking reported here (mostly due to the non-synchronised nature of \verb!fdb-hammer!), they were partially captured in both DAOS and Lustre runs. We expect that a full reproduction of such contention would result in lower Lustre bandwidths relative to DAOS, as the locking overheads would become more prominent.

For the reported \textit{no w+r contention} long runs with the DAOS and POSIX backends, \verb!fdb-hammer! was also run in its \verb!list()! mode to compare the listing performance of both backends. \verb!fdb-hammer! was configured to list all indexed fields for the first step archived in the write phase, and executed from a single node. Listing with the POSIX backend was consistently double as fast as with the DAOS backend, for all server node configurations. This is explained by the fact that the POSIX backend collocates all indexed field identifiers and locations for a same collocation key in a single file per process, and these files are loaded by the listing backend with a single read operation per file. In the DAOS backend, instead, every single indexed field location needs to be retrieved with a \texttt{daos\_kv\_get} operation, inflicting a large amount of I/O operations and causing strain on the DAOS server.

\section{Conclusion}

This work has demonstrated the feasibility of implementing Catalogue and Store backends for the FDB making use of DAOS, and presenting the same external semantics and API as the existing FDB backends. The performance of the new backends was tested, and we present a series of comparative scaling curves against our existing implementations running on top of a Lustre parallel file system on the same hardware.

Although implementing the same API, the new backend has some special characteristics. Most notably it immediately persists and makes data visible rather than waiting until explicit \verb|flush()| calls, and all data are indexed together rather than in one index per writing process. These should allow more aggressive workflow optimisation, and should support better scaling of read and listing semantics going forward. It is worth noting that obtaining optimal performance required adjusting the data schema, and thus the structure of the data indexing, between the backends.

The most relevant metrics obtained for operational NWP are the performance and scaling of runs with I/O patterns matching those in the forecast pipeline, with contention between simultaneously active readers and writers. Under these conditions, the DAOS backends have demonstrated higher throughput and better scaling than runs carried out using the POSIX backends on top of Lustre, on the same hardware. This argues strongly that the combination of the object storage semantics in DAOS, and the existence of the Key-Value objects in which contention is resolved locally on the DAOS server, give performance benefits over the mechanisms that can be built on top of distributed POSIX file systems.

The performance of the new DAOS backends tracks very closely to that of the Field I/O benchmark, which mimics the I/O approach used. This strongly confirms that the implemented backends are performing as expected, in line with the design aspirations.

The methodology applied in the performance analysis is well defined, and in particular provides good guidance on how to approach the parameter optimisation required when exploring such an extensive parameter domain. It is sufficiently general that it can be used to guide assessment of other I/O systems and developments.

We are thrilled by the outcome of this work, which has demonstrated the potential of DAOS for ECMWF and NWP workloads, and potentially for any other data-intensive workloads requiring relatively small object sizes or a high degree of contention. This opens new doors for ECMWF to consider a wider range of storage systems for operations in the future. In the immediate future we expect to test DAOS and the FDB backends on larger systems with an I/O configuration more closely matching actual operational use. Beyond this work, we are actively involved in research to enhance and extend high-performance workflows with semantically driven data movements and access, which this research will contribute to\cite{destine, warmworld, opencube, eupex}.

\begin{acks}
The work presented in this paper was carried out with funding by the European Union under the Destination Earth initiative (cost center DE3100, code 3320) and relates to tasks entrusted by the European Union to the European Centre for Medium-Range Weather Forecasts. Views and opinions expressed are those of the author(s) only and do not necessarily reflect those of the European Union or the European Commission. Neither the European Union nor the European Commission can be held responsible for them.

This work has also been contributed to by the ACROSS and IO-SEA projects, funded by the European High-Performance Computing Joint Undertaking (JU) under Grant Agreements no. 955648 and no. 955811, respectively. The JU receives support from the European Union’s Horizon 2020 research and innovation programme and Italy, France, Czech Republic, Greece, Netherlands, Germany, Norway.

The NEXTGenIO system was funded by the European Union's Horizon 2020 Research and Innovation program under Grant Agreement no. 671951, and supported by EPCC, The University of Edinburgh. Adrian Jackson was supported by UK Research and Innovation under the EPSRC grant EP/T028351/1.

For the purpose of open access, Adrian Jackson has applied a Creative Commons Attribution (CC BY) licence to any Author Accepted Manuscript version arising from this submission.
\end{acks}

\bibliographystyle{ACM-Reference-Format}

\end{document}